\documentclass[DIV=12,a4paper,11pt]{iopart}

\usepackage{color}
\usepackage{graphicx}
\usepackage{psfrag}
\usepackage{times}
\usepackage{iopams}
\usepackage{textcomp}
\usepackage{dsfont}
\usepackage{bm}
\usepackage{mathrsfs}

\usepackage[colorlinks=false, pdfborder={0 0 0}]{hyperref}

\usepackage[T1]{fontenc}
\usepackage{textcomp} 
\usepackage[latin1]{inputenc}
\usepackage{fixmath} 
\usepackage[nospace]{cite} 
\usepackage{exscale}

\newcommand{\Mi}{\rmi} 
\newcommand{\Me}{\rme} 
\newcommand{\Mbf}[1]{\bi{#1}} 
\newcommand{\Mdiff}{\mathrm{d}} 
\newcommand{\Mlabel}[1]{\mathrm{#1}} 

\newcommand{\eqref}[1]{(\ref{#1})}

\newcommand{\dd}{\partial}

\newcommand{\pdiff}[2]{\frac{\dd#1}{\dd#2}}

\renewcommand{\vec}[1]{\Mbf{#1}}
\newcommand{\mf}{m_\Mlabel{F}}
\newcommand{\wrf}{\omega_\Mlabel{rf}}
\newcommand{\ORabi}{\Omega_\Mlabel{rf}}
\newcommand{\x}{\vec{x}}

\begin{document}
\frenchspacing

\title[The space atom laser]{The space atom laser: An isotropic source for ultra-cold atoms in microgravity}
\author{Matthias~Meister$^1$, Albert~Roura$^1$, Ernst~M.~Rasel$^2$ and~Wolfgang~P.~Schleich$^{1,3}$}
\address{$^1$ Institut f\"ur Quantenphysik and Center for Integrated Quantum Science and Technology~(IQ$^\mathrm{ST}$),
Universit\"at Ulm, D-89081 Ulm, Germany}

\address{$^2$ Institut f\"ur Quantenoptik, Leibniz Universit\"at Hannover, D-30167 Hannover, Germany}

\address{$^3$ Hagler Institute for Advanced Study at Texas A\&M University, Texas A\&M AgriLife Research, Institute for Quantum Studies and Engineering (IQSE) and Department of Physics and Astronomy, Texas A\&M University, College Station, TX 77843-4242, USA}

\ead{matthias.meister@uni-ulm.de}

\begin{abstract}
Atom laser experiments with Bose-Einstein condensates (BECs) performed in ground-based laboratories feature a coherent and directed beam of atoms which is accelerated by gravity. 
In microgravity the situation is fundamentally different because the dynamics is entirely determined by the repulsive interaction between the atoms and not by the gravitational force. 
As a result, the output of a space atom laser is a spherical wave slowly expanding away from the initial BEC.
We present a thorough theoretical study of this new source of matter waves based on rf outcoupling which exhibits an isotropic distribution both in position and momentum even for an initially anisotropic trap.
The unique geometry of such a freely expanding, shell-shaped BEC offers new possibilities for matter waves in microgravity and is complementary to other matter-wave sources prepared by delta-kick collimation or adiabatic expansion.
Our work paves the way for the upcoming experimental realization of a space atom laser making use of NASA's Cold Atom Laboratory on the International Space Station.

\vspace{6cm}
\begin{center}
Article published in \textit{New Journal of Physics} \textbf{21}, 013039 (2019) 
\\
\url{https://doi.org/10.1088/1367-2630/aaf7b5}
\end{center} 

\end{abstract}

\maketitle

\section{Introduction}

Shortly after the first creation of a Bose-Einstein condensate (BEC) in a dilute gas \cite{Anderson1995,Davis1995} atom lasers based on this new state of matter have been demonstrated experimentally \cite{Mewes1997,Anderson1998,Hagley1999,Bloch1999} and described theoretically \cite{Ballagh1997,Naraschewski1997,Steck1997,Kneer1998,Kramer2002,Bracher2003,Kramer2006,Johnsson2007}. 
Since then atom lasers have been studied in a broad variety of experiments \cite{Robins2013} ranging from tests of their spatial \cite{Bloch2000} and temporal \cite{Koehl2001} coherence, the study of beam profiles \cite{Koehl2005,Riou2006} and the demonstration of Bragg reflectors \cite{Fabre2011} and beam splitters \cite{Gattobigio2012} to the improvement of their brightness \cite{Bolpasi2014}. 
All these experiments have been performed in earth-bound laboratories where the dominant force acting on the outcoupled atoms is gravity, which results in a directed and accelerated beam of atoms leaving the condensate. 
This acceleration can be reduced by applying a gravity-compensating potential with a magnetic field gradient \cite{KleineBuening2010} or by loading atom lasers into horizontal matter-wave guides \cite{Guerin2006,Couvert2008,Bernard2011}, but both approaches still yield a directed beam. 
Motivated by pioneering experiments with BECs in microgravity environments \cite{VanZoest2010,Muentinga2013,Becker2018}, we propose to enter a completely new regime for the atom laser by performing an experiment aboard the International Space Station (ISS), where the microgravity conditions allow us to create a unique shell-like atom laser which slowly expands away from the initial BEC driven only by the repulsive interaction between the atoms. This effect can be exploited to establish a nearly isotropic source for ultra-cold atoms in microgravity. 

In general, the formation of an atom laser is realized by coherently transferring a fraction of a magnetically trapped BEC to a magnetic insensitive state either via radio-frequency (rf) fields \cite{Mewes1997,Bloch1999,Martin1999,LeCoq2001}, microwave fields \cite{Oettl2005,Oettl2006} or Raman lasers \cite{Hagley1999,Robins2006,Debs2009}. 
In the absence of gravity the outcoupled atoms gain kinetic energy only through the repulsive interaction between the atoms, and thus expand away from the center of the trap. Consequently, the momentum distribution of the output state crucially depends on the location of the resonantly coupled shell during the outcoupling process. 
This relation has already been used in guided atom lasers to generate slowly propagating, directed beams of atoms for applications in guided atom optics \cite{Billy2007}.

In this paper we present a scheme based on rf coupling that enables the generation of a slowly expanding shell of atoms featuring an isotropic momentum distribution even if the initial BEC was trapped in an elongated anisotropic trap. 
For this purpose, we outcouple resonantly at the very edge of the BEC with relatively small coupling strength in order to establish a spatially well-localized and state-selective outcoupling process. In our case rf outcoupling is the preferred choice because in contrast to Raman lasers the atoms only get a negligible recoil kick, which is essential to obtain a spherically symmetric output. 
Moreover, it is experimentally less demanding to perform frequency ramps of a few Hertz with rf fields rather than microwave fields due to the much lower offset frequency of rf transitions. Finally, the drawback of driving a multi-level system inherent to rf outcoupling can be limited by making use of the asymmetric level structure generated by the Zeeman effect \cite{Bernard2011}.  
 
The outcoupling scheme presented in this paper has been developed for a direct implementation in future microgravity experiments like NASA's Cold Atom Laboratory (CAL) \cite{Elliott2018}, which will perform BEC experiments aboard the ISS. Thus, we have taken much care to incorporate all relevant experimental effects into our model and to choose realistic values for all parameters. In addition, we have also studied the influence of  fluctuations of the initial number of particles and the offset magnetic field on the performance of the space atom laser. Hence, this paper is not only a theoretical study of rf outcoupling in microgravity but it is also intended to pave the way for an experimental realization of the space atom laser in the near future. 

Our paper is structured as follows.
In section~\ref{sec:theoretical:basics} we lay out the theoretical framework for the description of the space atom laser based on coupled Gross-Pitaevskii (GP) equations and derive conditions for a spatially-localized and state-selective outcoupling. In addition, we discuss the origin of the unique isotropy of the output state of the space atom laser and why this isotropy is essentially independent of the geometry of the magnetic trap. 

A model sequence that allows to generate such an isotropic output state is presented in section~\ref{sec:sample:test:sequence} together with the corresponding results obtained by numerical simulations.  
The shape of the outcoupled shell is discussed in more detail in section~\ref{sec:isotropy}, where we compare the isotropy of the momentum distribution for different coupling strengths. Furthermore, we show how second-order contributions to the Zeeman effect lead to a slight broadening of the momentum distribution.

The influence of varying experimental parameters on the output of the space atom laser is studied in section~\ref{sec:parameter:scans} by numerically scanning through the initial number of particles and the Rabi frequency of the rf field. In this way we identify different regimes of operation for the space atom laser including a fairly stable regime with large isotropy for intermediate coupling as well as an oscillatory regime for strong coupling. Finally, we have analyzed the impact of fluctuations of the offset magnetic field on the rf-outcoupling process with additional numerical simulations. As it turns out, even small changes of the magnetic field can shift the resonance of the rf field far away from the BEC so that the outcoupling shuts down completely. Hence, getting the magnetic field fluctuations under control is crucial for a successful experimental realization of the space atom laser.

In section~\ref{sec:conclusion} we briefly summarize our results and give an outlook towards future experiments and applications. 
Some of the more technical calculations are carried out in two appendices that conclude this paper. In \ref{app:sec:general:zeeman:effect} we discuss the Zeeman effect and the Breit-Rabi formula used to compute the magnetic trapping potentials and derive an approximate relation for the anti-trapping potential of the $\mf=0$ hyperfine sublevel. In \ref{app:sec:consitions:sharp:and:state} we provide a more detailed derivation of the conditions for a spatially-dependent and state-selective outcoupling.

\section{Theoretical basis}\label{sec:theoretical:basics}
Throughout this article we consider a magnetically trapped BEC of $^{87}$Rb atoms which are in the electronic ground state and within the hyperfine manifold with total angular momentum $F=1$. Nevertheless, the ideas and concepts presented here can be straightforwardly generalized to other suitable isotopes like $^{23}$Na, $^{39}$K and $^{41}$K as well as to the case $F = 2$.

In this section we present the theoretical framework for the description of an atom laser in microgravity. We first discuss the coupled GP equations that determine the dynamics of the different hyperfine levels and then derive two conditions that enable us to perform a spatially-localized as well as state-selective outcoupling. Furthermore, we argue why the output of the space atom laser has an isotropic distribution in position and momentum space. This section concludes with a brief summary of the coupled GP equations in the rotating-wave approximation.

\subsection{Coupled Gross-Pitaevskii equations}
In the presence of a magnetic field the $F=1$ ground-state hyperfine manifold of $^{87}$Rb is split into a trapped ($\mf = -1$), untrapped ($\mf = 0$) and anti-trapped ($\mf = +1$) sublevel, which are respectively described by the corresponding wave functions $\psi_{\mf} = \psi_{\mf}(\x,t)$. Each of these wave functions fulfills the normalization condition
\begin{equation}\label{eq:normalization}
 \int \Mdiff^3 x \,\left|\psi_{\mf}(\x,t)\right|^2 = N_{\mf} \;,
\end{equation}
where $N_{\mf}$ is the number of particles in that sublevel. 

When a monochromatic rf field of frequency $\wrf$ is applied to drive transitions between neighboring sublevels, the dynamics is governed by the following system of three coupled GP equations \cite{Ballagh1997,Steck1997}:
\begin{equation}\label{eq:coupled:GPE}
\Mi\hbar \,\pdiff{}{t}\left( \begin{array}{l}
\psi_{-1} \\ \psi_{0} \\ \psi_{+1}
\end{array} \right)
= \left(\begin{array}{ccccc}
H_{-1}  && H_\mathrm{rf} && 0 \\
H_\mathrm{rf} && H_{0} && H_\mathrm{rf} \\
0 && H_\mathrm{rf} && H_{+1}
\end{array}\right)
\left( \begin{array}{l}
\psi_{-1} \\ \psi_{0} \\ \psi_{+1}
\end{array} \right) \;.
\end{equation}
The diagonal elements of the matrix in Eq. \eqref{eq:coupled:GPE} are determined by the usual GP Hamiltonian 
\begin{equation}\label{eq:Hamiltonian:mF}
 H_{\mf} = H_{\mf}(\x,t) = -\frac{\hbar^2 \nabla_\x^2}{2M} + V_{\mathrm{ext},\mf}(\x,t) + V_{\mathrm{int},\mf}(\x,t) \;.
\end{equation} 
Here, $M$ denotes the mass of a single atom and $V_{\mathrm{ext},\mf}(\x,t)$ is the sublevel-dependent external trapping potential 
\begin{equation}\label{eq:external:potential}
 V_{\mathrm{ext},\mf}(\x,t) = V_{\mathrm{bot},\mf} + V_{\mathrm{trap},\mf}(\x,t) \;,
\end{equation}
which consists of the constant offset potential $V_{\mathrm{bot},\mf}$ corresponding to the trap bottom and the spatially dependent part $V_{\mathrm{trap},\mf}(\x,t)$ that actually defines the geometry of the trap. Furthermore, $V_{\mathrm{int},\mf}(\x,t)$ is the mean-field interaction potential defined by
\begin{equation}\label{eq:interaction:potential}
 V_{\mathrm{int},\mf}(\x,t) = \sum\limits_{{\mf}'} g_{{\mf},{\mf}'} \left|\psi_{{\mf}'}(\x,t) \right|^2 \;,
\end{equation}
where $g_{{\mf},{\mf}'} = 4\pi\hbar^2 a_{{\mf},{\mf}'} / M$ is the coupling constant between the two sublevels $\mf$ and ${\mf}'$ with the s-wave scattering length $a_{{\mf},{\mf}'}$.

The off-diagonal elements of the Hamiltonian matrix on the right-hand side of Eq. \eqref{eq:coupled:GPE} are responsible for the transitions between neighboring hyperfine sublevels driven by the rf field. Within the magnetic dipole approximation the corresponding matrix element $H_\mathrm{rf}$ is given by the relation
\begin{equation}\label{eq:Hamiltonian:rf}
 H_\mathrm{rf} = H_\mathrm{rf}(t) = \hbar\ORabi\, \cos(\wrf\;t) 
\end{equation}
with the Rabi frequency
\begin{equation}\label{eq:Omega:Rabi}
 \ORabi = \frac{|g_\mathrm{F}| \,\mu_\mathrm{B} \,B_\mathrm{rf}}{\hbar} \;,
\end{equation}
which depends on the hyperfine Land\'e $g$-factor $g_\mathrm{F}$, the Bohr magneton $\mu_\mathrm{B}$ and the amplitude $B_\mathrm{rf}$ of the rf field. Throughout this paper we neglect any spatial dependence of the strength and the orientation of the rf field and assume that $\ORabi$ is constant across the coupling region. 

Initially, the BEC is prepared in the ground state of the trapped magnetic sublevel, $\mf=-1$, whose wave function is given by the lowest-energy solution of the time-independent GP equation
\begin{equation}\label{eq:time:independent:GP}
 \mu_{-1}\,\psi_{-1}(\x,0) = \left[-\frac{\hbar^2 \nabla_\x^2}{2M} + V_{\mathrm{trap},-1}(\x,0) + V_{\mathrm{int},-1}(\x,0)\right] \,\psi_{-1}(\x,0) \,,
\end{equation}
where $\mu_{-1}$ is the chemical potential.

\subsection{Spatially-dependent, state-selective outcoupling}\label{subsec:spatial-dependent:st4ate:selective}
In order to efficiently outcouple a significant number of particles from the trapped BEC to the untrapped state forming a slowly expanding shell of atoms, it is essential (i) to resonantly outcouple only at the edge of the condensate to reduce the amount of kinetic energy that the transferred particles gain through the repulsive interaction and (ii) to suppress the outcoupling to the anti-trapped state, in which the atoms are strongly accelerated. In this subsection we discuss how these two conditions can be satisfied during the outcoupling process. 

To highlight the relevant features of the rf outcoupling, we consider a slightly simplified model of the BEC dynamics by (i) neglecting the kinetic energy term in the GP Hamiltonian~\eqref{eq:Hamiltonian:mF} in the spirit of the Thomas-Fermi approximation \cite{Dalfovo1999} and by (ii) assuming that the coupling constants between all hyperfine sublevels are identical, that is $g_{{\mf},{\mf}'} \approx g$, which for $^{87}$Rb is a very accurate approximation \cite{StamperKurn2013}. The interaction potential \eqref{eq:interaction:potential} then reduces to
\begin{equation}\label{eq:equal:interaction:potential}
 V_{\mathrm{int},\mf}(\x,t) \approx V_{\mathrm{int}}(\x,t) = g \sum\limits_{{\mf}'} \left|\psi_{{\mf}'}(\x,t) \right|^2 \;
\end{equation}
and is identical for all sublevels.
We stress that these two assumptions are only used in the context of this section to illustrate how a spatially-localized, state-selective outcoupling can in principle be realized. In later sections we will use again the full GP Hamiltonian \eqref{eq:Hamiltonian:mF}.

Having the approximations mentioned above in mind, we define the potential energy of the hyperfine states by
\begin{equation}\label{eq:energy:levels}
 E_{\mf} = E_{\mf}(\x,t) = V_{\mathrm{ext},\mf}(\x,t) + V_{\mathrm{int}}(\x,t) \,.
\end{equation}
The resonant transition frequency $\omega_{\mf,{\mf}'}$ between two sublevels is given by the difference of the corresponding potential energies
\begin{equation}\label{eq:transition:frequency}
 \omega_{\mf,{\mf}'} = \left| E_{\mf} - E_{{\mf}'}\right| /\hbar \,. 
\end{equation}
Hence, the relevant transition frequencies for our model 
\begin{eqnarray}
 \omega_{-1,0} = \omega_{-1,0}(\x,t) = \left[V_{\mathrm{ext},-1}(\x,t) - V_{\mathrm{ext},0}(\x,t)\right]/\hbar \,, \label{eq:reso:frequency:-1:0} \\
 \omega_{0,+1} = \omega_{0,+1}(\x,t) = \left[V_{\mathrm{ext},0}(\x,t) - V_{\mathrm{ext},+1}(\x,t)\right]/\hbar   \label{eq:reso:frequency:0:+1}
\end{eqnarray}
are entirely determined by the external trapping potentials and are, therefore, spatially dependent in general. 

Since we focus on purely magnetic traps, the potential $V_{\mathrm{ext},\mf}(\x,t)$ is determined by the magnetic field $\vec{B}(\x,t)$. For convenience we split the absolute value of $\vec{B}(\x,t)$ according to the relation 
\begin{equation}\label{eq:magnetic:field:split}
 |\vec{B}(\x,t)| = B_{\mathrm{bot}} + B_{\mathrm{trap}}(\x,t) \;,
\end{equation}
where $B_{\mathrm{bot}}$ corresponds to the constant offset field at the trap bottom and $B_{\mathrm{trap}}(\x,t)$ generates the spatially dependent trapping potential. The quantization axis for the hyperfine sublevels is determined mainly by the orientation of the offset magnetic field $B_{\mathrm{bot}}$.
If one considers the \emph{linear Zeeman effect}, characterized by 
\begin{equation}\label{eq:lin:Zeeman:effect}
 V_{\mathrm{ext},\mf}(\x,t) = \mf \,g_\mathrm{F} \, \mu_\mathrm{B} |\vec{B}(\x,t)| \;,
\end{equation}
one might be tempted to assume that the transition frequencies \eqref{eq:reso:frequency:-1:0} and \eqref{eq:reso:frequency:0:+1} are identical. 

However, even for relatively low magnetic fields of a few Gauss, the \emph{general Zeeman effect} \cite{BergmannSchaefer1997} leads to an asymmetry between the potential energies of the trapped and anti-trapped states which is of the order of the chemical potential of the initial ground state defined by Eq. \eqref{eq:time:independent:GP}. Thus, it needs to be taken into account when determining the transition frequencies \eqref{eq:reso:frequency:-1:0} and \eqref{eq:reso:frequency:0:+1}.
In \ref{app:sec:Breit_Rabi} we discuss the Breit-Rabi formula \cite{BreitRabi1931} for the Zeeman effect, which can be used to describe the correct energy splitting of the hyperfine states. The necessity of using the Breit-Rabi formula will become clear in section \ref{sec:sample:test:sequence}, when we discuss quantitatively the space atom laser and evaluate the relevant quantities, such as the chemical potential of the initial BEC and the transition frequencies. Hence, instead of applying Eq. \eqref{eq:lin:Zeeman:effect}, we make use of the Breit-Rabi formula 
\begin{equation}\label{eq:Breit-Rabi-formula_Rb87}
\fl V_{\mathrm{ext},\mf}(\x,t) = - \frac{A_\mathrm{hfs}}{4} - \mf\, g_I\, \mu_\mathrm{B}\, \big|\vec{B}(\x,t)\big| - A_\mathrm{hfs} \sqrt{1 + \mf\,b(\x,t) + b^2(\x,t) }
\end{equation}
with
\begin{equation}\label{eq:def:b}
 b(\x,t) = \frac{(g_J - g_I)\, \mu_\mathrm{B}\, \big|\vec{B}(\x,t)\big|}{2 A_{\mathrm{hfs}}} \;,
\end{equation}
which is valid for the $F=1$ ground-state manifold of $^{87}$Rb and enables us to calculate precisely the trapping potentials generated by the magnetic field $\vec{B}(\x,t)$. Here, $A_{\mathrm{hfs}}$ is the zero-field hyperfine constant of the atom, $g_I$ and $g_J$ are respectively the nuclear and the fine-structure $g$-factors. 

As shown in \ref{app:sec:Breit_Rabi}, by applying the Breit-Rabi formula \eqref{eq:Breit-Rabi-formula_Rb87}, we obtain the relation
\begin{equation}\label{eq:transition:frequencies:not:equal}
 \omega_{-1,0} > \omega_{0,+1} 
\end{equation}
for the transition frequencies \eqref{eq:reso:frequency:-1:0} and \eqref{eq:reso:frequency:0:+1}, which implies an asymmetry between the potential energies of the trapped and the anti-trapped states. 

In addition, according to the Breit-Rabi formula in Eq.~\eqref{eq:Breit-Rabi-formula_Rb87} the trapping potential for the $\mf = 0$ state actually exhibits a spatial dependence which, as derived in \ref{app:sec:mf0_pot}, is given by the relation
\begin{equation}\label{eq:potential:0:spatial:dependence}
 V_{\mathrm{trap},0}(\x,t) \approx - \frac{(g_J - g_I)^2 \,\mu_\mathrm{B}}{4\,|g_F| \,A_\mathrm{hfs}} \,B_\mathrm{bot} \, V_{\mathrm{trap},-1}(\x,t)
\end{equation}
valid for $b^2(\x,t) \ll 1$. Hence, atoms in the $\mf = 0$ state are subject to a small anti-trapping potential, which is proportional to $- V_{\mathrm{trap},-1}(\x,t)$ and accelerates them away from the center of the trap. In order to suppress the effect of this additional potential, the magnetic-field offset $B_\mathrm{bot}$ needs to fulfill the condition
\begin{equation}\label{eq:condition:small:anti:trapping:mF0}
 B_\mathrm{bot} \ll \frac{4\,|g_F| \,A_\mathrm{hfs}}{(g_J - g_I)^2 \,\mu_\mathrm{B}} \;.
\end{equation}

\begin{figure}[h]
\begin{center}
 \includegraphics[scale=0.8]{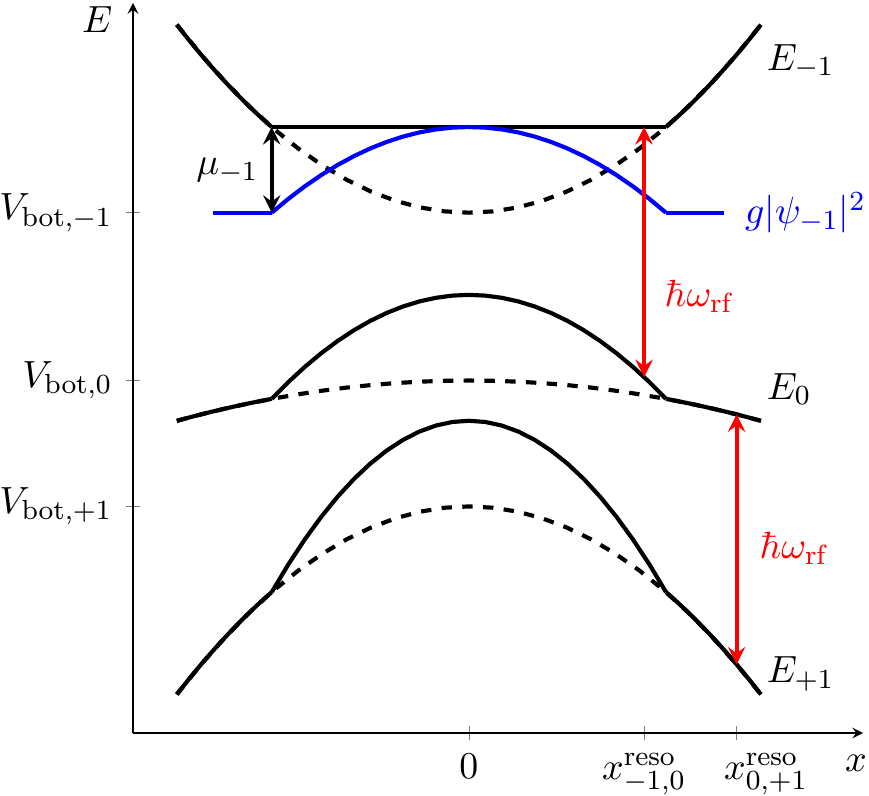}
 \caption{Sketch of the potential energies $E_{\mf}$ of the $\mf = -1, 0$ and $+1$ hyperfine states of a BEC in a harmonic magnetic trap, where all the atoms occupy the trapped state $\mf = -1$. The dashed lines indicate the external potential $V_{\mathrm{ext},\mf}$, while the blue line depicts the interaction potential $V_{\mathrm{int},\mf}$. The black solid lines show the sum of these two potentials for each hyperfine state. The energy axis is not to scale but displays the asymmetry between the  $\mf = -1$ and $+1$ state as well as the small anti-trapping potential for the $\mf=0$ state. The red arrows illustrate that for a given rf frequency $\wrf$ the resonance conditions \eqref{eq:reso:condition:-1:0} and \eqref{eq:reso:condition:0:+1} are fulfilled at different positions.}
 \label{fig:energy_levels}
\end{center}
\end{figure}

A schematic overview of the potential energies \eqref{eq:energy:levels} taking into account the asymmetry \eqref{eq:transition:frequencies:not:equal} and the small anti-trapping potential \eqref{eq:potential:0:spatial:dependence} for the $\mf=0$ state is displayed in Fig.~\ref{fig:energy_levels} for the example of a BEC in a harmonic trap. 
As one can clearly see, the energy gap between neighboring sublevels strongly varies in space so that the position $\x^{\mathrm{reso}}_{-1,0}(t)$ where the resonance condition
\begin{equation}\label{eq:reso:condition:-1:0}
 \omega_{-1,0}\left(\x^{\mathrm{reso}}_{-1,0}(t),t\right) = \wrf(t)
\end{equation}
to transfer atoms from the trapped to the untrapped state is fulfilled can be tuned easily by changing the rf frequency $\wrf(t)$ appropriately. The same statement holds true for the resonance condition 
\begin{equation}\label{eq:reso:condition:0:+1}
 \omega_{0,+1}\left(\x^{\mathrm{reso}}_{0,+1}(t),t\right) = \wrf(t)
\end{equation}
for the transition from the untrapped to the anti-trapped state.

In addition, Fig.~\ref{fig:energy_levels} shows that the asymmetry between the two transition frequencies
ensures that for a single rf frequency $\wrf$ the resonance position $\x^{\mathrm{reso}}_{0,+1}$ is shifted further away from the center of the trap compared with the resonance position $\x^{\mathrm{reso}}_{-1,0}$. As a consequence, atoms outcoupled from the trapped to the untrapped state will not be immediately transferred to the anti-trapped state, which allows to transfer selectively the atoms from one state to another.

Hence, we have shown that by controlling the rf frequency $\wrf(t)$ appropriately, it is possible to achieve a spatially-dependent as well as state-selective outcoupling. 
For the remainder of this paper, we split the rf frequency
\begin{equation}\label{eq:rf:frequency:split}
 \omega_\Mlabel{rf}(t) = \omega_0 + \Delta_\Mlabel{rf}(t) 
\end{equation}
into two parts, where the constant offset frequency 
\begin{equation}\label{eq:def:omega_0}
 \omega_0 = \left[ V_{\mathrm{bot},-1} - V_{\mathrm{bot},0} \right]/\hbar
\end{equation}
accounts for the main energy difference between the trapped and the untrapped state. The time dependent term $\Delta_\Mlabel{rf}(t)$ determines the resonance position $\x^{\mathrm{reso}}_{-1,0}(t)$ according to the relation
\begin{equation}\label{eq:reso:Drf}
 \left[ V_{\mathrm{trap},-1}\left(\x^{\mathrm{reso}}_{-1,0}(t),t\right) - V_{\mathrm{trap},0}\left(\x^{\mathrm{reso}}_{-1,0}(t),t\right) \right]/\hbar = \Delta_\Mlabel{rf}(t) \,,
\end{equation}
which directly follows from Eq. \eqref{eq:reso:condition:-1:0} by applying Eqs. \eqref{eq:reso:frequency:-1:0}, \eqref{eq:rf:frequency:split} and \eqref{eq:def:omega_0}.

Finally, in order to precisely address atoms only at the edge of the BEC and to avoid transferring them to the anti-trapped state, it is crucial to have a spatially well-localized resonance since otherwise the whole BEC would be outcoupled at once. As discussed in \ref{app:sharp:resonance} such a sharp resonance requires the condition
\begin{equation}\label{eq:sharp:resonance:condition}
 \mu_{-1}^2 \gg (\hbar\ORabi)^2  
\end{equation}
to be fulfilled, which sets an upper bound on the Rabi frequency.
Obviously, small values of $\ORabi$ allow for a better control of where the outcoupling process takes place and also reduce the amount of back-coupling to the initial state, while a large coupling strength leads to the emergence of full Rabi oscillations between neighboring states.

Additionally, the asymmetry between the two transition frequencies $\omega_{-1,0}$ and $\omega_{0,+1}$ displayed in Eq. \eqref{eq:transition:frequencies:not:equal} can be increased by raising the magnetic field $B_\mathrm{bot}$ at the trap bottom. If the condition
\begin{equation}\label{eq:state:selective:condition}
 B_\mathrm{bot}^2 \gg \frac{16 \,\mu_{-1} \,A_\mathrm{hfs}}{(g_J - g_I)^2 \,\mu_\mathrm{B}^2}
\end{equation}
derived in \ref{app:state:selective} holds true, $\omega_{-1,0}$ and $\omega_{0,+1}$ are shifted away from each other far enough to suppress the coupling to the anti-trapped state nearly completely. Hence, by choosing $\ORabi$ and $B_\mathrm{bot}$ appropriately compared to the chemical potential $\mu_{-1}$ a spatially-localized and state-selective outcoupling can be realized. Unfortunately, according to Eq. \eqref{eq:potential:0:spatial:dependence} raising the magnetic-field offset $B_\mathrm{bot}$ also leads to an acceleration of the outcoupled atoms due to the small anti-trapping potential caused by the general Zeeman effect. Consequently, there is a trade-off between the efficiency of outcoupling atoms into the $\mf=0$ state and the resulting expansion velocity of these atoms.

\subsection{Output state}\label{subsec:output:state}
Based on the results of the previous section, we are now able to make qualitative predictions about the shape of the output state of the space atom laser. 
Under the assumption that the outcoupling process takes only place at the edge of the BEC and is slow enough not to excite collective oscillations of the remaining trapped atoms, we can treat $\psi_{-1}(\x,t)$ as being in the ground state determined by the time-independent GP equation~\eqref{eq:time:independent:GP}, but with a decreasing number of particles. When in addition we neglect the kinetic energy term according to the Thomas-Fermi approximation \cite{Dalfovo1999}, and make use of Eq. \eqref{eq:equal:interaction:potential}, we can define the time-dependent chemical potential
\begin{equation}\label{eq:time:dependent:chemical:potential}
 \mu_{-1}(t) \approx V_{\mathrm{trap},-1}(\x,t) + V_{\mathrm{int}}(\x,t)
\end{equation}
valid within the trapped BEC. We stress that this approximation oversimplifies the true dynamics of the outcoupling process, especially close to the edge of the BEC, but it is still useful to get an intuitive understanding of the general properties of the outcoupled state. 

By combining Eqs. \eqref{eq:potential:0:spatial:dependence} and \eqref{eq:reso:Drf} we obtain the relation
\begin{equation}\label{eq:potential:-1:at:resonance}
 V_{\mathrm{trap},-1}\left(\x^{\mathrm{reso}}_{-1,0}(t),t\right) \approx \frac{\hbar\,\Delta_\Mlabel{rf}(t)}{1 + \frac{(g_J-g_I)^2 \,\mu_\mathrm{B}\,B_\mathrm{bot}}{4\,|g_F|\,A_\mathrm{hfs}}} \;,
\end{equation}
which states that the trapping potential for the $\mf=-1$ state is spatially constant over the whole resonance area. 

Applying Eq. \eqref{eq:potential:-1:at:resonance} to Eq. \eqref{eq:time:dependent:chemical:potential} yields a similar relation for the interaction potential
\begin{equation}\label{eq:interaction:potential:-1:resonance}
 V_{\mathrm{int}}(\x^{\mathrm{reso}}_{-1,0}(t),t) \approx \mu_{-1}(t) - \frac{\hbar\,\Delta_\Mlabel{rf}(t)}{1 + \frac{(g_J-g_I)^2 \,\mu_\mathrm{B}\,B_\mathrm{bot}}{4\,|g_F|\,A_\mathrm{hfs}}} \;.
\end{equation}
Hence, the interaction energy of the outcoupled atoms is approximately equal across the resonance area, and consequently the amount of kinetic energy that these atoms gain by running down the interaction hill is equal, too. As a consequence, the momentum distribution of the space atom laser is a thin spherical shell centered around zero. We emphasize that the geometry of the initially trapped BEC does not play any role in this derivation. Thus, even when the initial BEC is trapped in an elongated, anisotropic trap, the momentum distribution of the outcoupled shell has a spherical shape. 

However, the distribution of the momentum density within this spherical shell is in general not isotropic. Indeed, its angular dependence is determined by the coupling strength of the rf field as well as by the shape of the trapped BEC as we will show in section~\ref{sec:isotropy}. Finally, after a sufficiently long free evolution time the density distribution in position space is determined by the momentum distribution and therefore also has the shape of a spherical shell. 

For a more comprehensive study of the output state of the space atom laser several additional effects need to be considered that lead to deviations from the ideal spherical shape of the momentum distribution: (i) collective excitations of the trapped BEC result in non-vanishing momenta of the just outcoupled atoms and (ii) the small anti-trapping potential for the $\mf=0$ state, Eq. \eqref{eq:potential:0:spatial:dependence}, introduces a spatially dependent acceleration. Both of these effects depend on the geometry of the magnetic trap and are, thus, highly anisotropic in general. Nevertheless, if these effects can be kept small, the output of the space atom laser is quite isotropic both in position and momentum space.

\subsection{Rotating-wave approximation}
In order to eliminate the oscillating terms in Eq.~\eqref{eq:coupled:GPE}, which is necessary to perform analytical calculations or numerical simulations of the dynamics, we transform to the internal rotating frame by introducing the wave functions 
\begin{equation}
\tilde\psi_{\mf} = \Me^{-\Mi\,(\mf\,\omega_0 - V_{\mathrm{bot},0} / \hbar)\,t} \,\psi_{\mf} .
\end{equation}
As long as $|\omega_0| \gg |\Delta_\Mlabel{rf}(t)|$ for all times $t$, we can safely apply the rotating-wave approximation \cite{Schleich2001} and rewrite Eq.~\eqref{eq:coupled:GPE} accordingly to obtain the new set of equations
\begin{equation}\label{eq:coupled:GPE:rot:wave:approx}
\fl \Mi\hbar \,\pdiff{}{t}\left( \begin{array}{l}
\tilde\psi_{-1} \\ \tilde\psi_{0} \\ \tilde\psi_{+1}
\end{array} \right)
= \left(\begin{array}{ccccc}
\tilde{H}_{-1} && \frac{1}{2}\hbar\ORabi \,\Me^{-\Mi\Delta_\Mlabel{rf}\,t} && 0 \\
\frac{1}{2}\hbar\ORabi \,\Me^{+\Mi\Delta_\Mlabel{rf}\,t} && \tilde{H}_{0} && \frac{1}{2}\hbar\ORabi \,\Me^{-\Mi\Delta_\Mlabel{rf}\,t} \\
0 && \frac{1}{2}\hbar\ORabi \,\Me^{+\Mi\Delta_\Mlabel{rf}\,t} && \tilde{H}_{+1} + V_\mathrm{BR}
\end{array}\right)
\left( \begin{array}{l}
\tilde\psi_{-1} \\ \tilde\psi_{0} \\ \tilde\psi_{+1}
\end{array} \right) \;,
\end{equation}
where the Hamiltonian $\tilde{H}_{\mf}$ on the diagonal is defined by
\begin{equation}
\fl \tilde{H}_{\mf} = \tilde{H}_{\mf}(\x,t) = H_{\mf}(\x,t) - V_{\mathrm{bot},\mf} = -\frac{\hbar^2 \nabla_\x^2}{2M} + V_{\mathrm{trap},\mf}(\x,t) + V_{\mathrm{int},\mf}(\x,t)\;.
\end{equation}
The additional term
\begin{equation}
 V_\mathrm{BR} = V_{\mathrm{bot},-1} + V_{\mathrm{bot},+1} - 2 V_{\mathrm{bot},0}
\end{equation}
in Eq. \eqref{eq:coupled:GPE:rot:wave:approx} accounts for the asymmetric splitting of the hyperfine sublevels due to the Zeeman effect calculated with the Breit-Rabi formula \eqref{eq:Breit-Rabi-formula_Rb87}.

When comparing Eqs.~\eqref{eq:coupled:GPE} and \eqref{eq:coupled:GPE:rot:wave:approx}, we find that the amount of constant offset terms on the diagonal of the matrix as well as the frequency of the time-dependent oscillations on the off-diagonal elements has been reduced significantly. Thus, we can now use this set of differential equations to determine the dynamics of the space atom laser. 
In order to further reduce the computational costs of solving Eq.~\eqref{eq:coupled:GPE:rot:wave:approx} the transformations and scalings described in Refs.~\cite{Nandi2007,Meister2017} can be applied, which are especially beneficial for long free evolution times of the output state.

\section{Model sequence}\label{sec:sample:test:sequence}
After having discussed the theoretical basis of the space atom laser we present in this section the numerical results of a model sequence that illustrate what to expect from actually performing an rf outcoupling experiment with BECs in microgravity. We first describe the setup of our test sequence and show that the parameters fulfill the conditions for a spatially-dependent and state-selective outcoupling as described in section \ref{subsec:spatial-dependent:st4ate:selective}. We then present the results of our numerical simulations and analyse the characteristic effects of the space atom laser.

\subsection{Setup}\label{subsec:setup}
The setup of our model sequence is based on realistic values that are in accordance with the planned capabilities of the CAL apparatus \cite{CAL_Homepage}. Table~\ref{tab:parameters:sample:sequence} contains the values of the relevant physical quantities, which correspond to only one of the many different scenarios that we have been testing for our study. The influence of different parameters on the outcoupling result is discussed in section~\ref{sec:parameter:scans}.
\begin{table}[h]
\begin{center}
\caption{List of the physical parameters for the model sequence.}
\begin{tabular}{llcl}
\\ \hline\hline \\
number of particles & $N$ & $=$ & $10^5$ \\
trap frequencies & $\omega_x = \omega_y$ & $=$ & $2\pi \cdot 30\, \mathrm{Hz}$ \\
 & $\omega_z$ & $=$ & $2\pi \cdot 15\, \mathrm{Hz}$ \\
magnetic field minimum & $B_\mathrm{bot}$ & $=$ & $4\, \mathrm{G}$ \\
Rabi frequency & $\ORabi$ & $=$ & $2\pi \cdot 90\,\mathrm{Hz}$ \\
rf offset & $\omega_0$ & $=$ & $2\pi \cdot 2.799\,\mathrm{MHz}$ \\
rf-outcoupling time & $t_\mathrm{rf}$ & $=$ & $90\, \mathrm{ms}$ \\
full evolution time & $t_\mathrm{max}$ & $=$ & $140\, \mathrm{ms}$ \\
\\ \hline\hline 
\end{tabular}
\label{tab:parameters:sample:sequence}
\end{center}
\end{table}

For the sake of simplicity we have chosen a cylindrically symmetric trap, where the trapping potentials for the different hyperfine sublevels are calculated with the Breit-Rabi formula \eqref{eq:Breit-Rabi-formula_Rb87} and the harmonic magnetic field
\begin{equation}\label{eq:def:harmonic:magnetic:field}
 |\vec{B}(\x,t)| = B_\mathrm{bot} + \frac{\partial^2|\vec{B}(\x,t)|}{\partial x^2}\,\frac{x^2}{2} + \frac{\partial^2|\vec{B}(\x,t)|}{\partial y^2}\,\frac{y^2}{2} + \frac{\partial^2|\vec{B}(\x,t)|}{\partial z^2}\,\frac{z^2}{2}  \;.
\end{equation}
The curvatures of the magnetic field have been chosen to generate in first order a harmonic trap for the $\mf=-1$ state with the trap frequencies $\omega_x$, $\omega_y$ and $\omega_z$. In order to do so, we have used the linear Zeeman effect, Eq. \eqref{eq:lin:Zeeman:effect}, leading to the relation
\begin{equation}
\frac{\partial^2|\vec{B}(\x,t)|}{\partial x_i^2} = \frac{M\,\omega_i^2}{|g_F|\, \mu_\mathrm{B}} \qquad \mathrm{with} \quad i = x,y,z \;.
\end{equation}

\begin{figure}[h]
\begin{center}
 \includegraphics[scale=0.9]{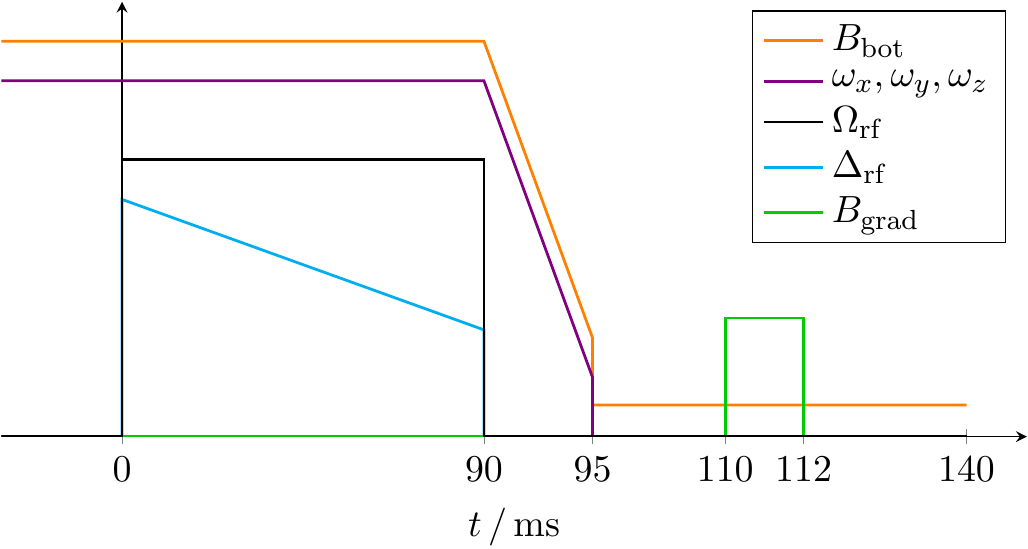}
 \caption{Schematic diagram of the outcoupling sequence featuring a linear ramp-down of the rf detuning $\Delta_\Mlabel{rf}(t)$, a subsequent ramp-down of the offset magnetic field $B_\mathrm{bot}$ and the trap frequencies $\omega_x$, $\omega_y$ and $\omega_z$ as well as a short pulse with a magnetic field $B_\mathrm{grad}$ exhibiting a linear gradient along the $z$-axis.}
 \label{fig:Drf:sequence}
\end{center}
\end{figure}

With the number of particles and the trap frequencies listed in Tab. \ref{tab:parameters:sample:sequence} we numerically obtain the chemical potential 
\begin{equation}\label{eq:chem:pot:numeric:result}
\mu_{-1}/\hbar = 2\pi \cdot 318 \,\mathrm{Hz}
\end{equation}
for the ground state of the system as the solution to the time-independent GP equation \eqref{eq:time:independent:GP}. This ground state represents the starting point of our outcoupling sequence, which is illustrated in Fig.~\ref{fig:Drf:sequence}. 

At the time $t=0$ the rf field with the Rabi frequency $\ORabi$ is switched on and the rf frequency $\omega_\mathrm{rf}(t)$, defined by Eq. \eqref{eq:rf:frequency:split}, is linearly ramped down over $90\,\mathrm{ms}$ according to the time dependent detuning
\begin{equation}\label{eq:Delta:rf:test:sequence}
 \Delta_{\mathrm{rf}}(t) = 2\pi \cdot \left(319 \,\mathrm{Hz} - 1.6 \;\frac{\mathrm{Hz}}{\mathrm{ms}} \cdot t \right) \qquad \mathrm{for} \quad 0 \leq t \leq 90\,\mathrm{ms} \,.
\end{equation}
At the end of this ramp-down the rf field is switched off completely.
Subsequently, the trap bottom as well as the trap frequencies are linearly ramped down over $5\,\mathrm{ms}$ from $B_\mathrm{bot} = 4\,\mathrm{G}$ to $1\,\mathrm{G}$ and from $\omega_i$ as listed in Tab.~\ref{tab:parameters:sample:sequence} to $0.1\, \omega_i$, respectively. At the time $t=95\,\mathrm{ms}$ the magnetic trap is switched off, while the offset magnetic field is reduced to $B_\mathrm{bot} = 0.2\,\mathrm{G}$. From this point onward the whole system evolves freely till $t=110\,\mathrm{ms}$, when the magnetic gradient field
\begin{equation}\label{eq:magnetic:field:gradient}
 B_\mathrm{grad}(\x,t) = 1 \,\frac{\mathrm{G}}{\mathrm{mm}} \cdot z \qquad \mathrm{for} \quad 110\,\mathrm{ms} \leq t \leq 112\,\mathrm{ms}
\end{equation}
is applied for $2\,\mathrm{ms}$. Finally, at $t=140\,\mathrm{ms}$ the simulation ends and the final states are analyzed. 

The time dependence of the rf frequency given by Eq. \eqref{eq:Delta:rf:test:sequence} ensures that the outcoupling process starts resonantly at the very edge of the BEC, and the resonance is then shifted closer and closer to the center of the trap as more and more atoms are transferred from the trapped to the untrapped state. In this way we can continuously empty the BEC from the edge to the center as proposed in Ref.~\cite{Johnsson2007}, while ensuring that the kinetic energy gained per particle stays constant during the whole outcoupling process. 

By keeping the magnetic trap switched on for a short time after the rf outcoupling has stopped, the remaining atoms in the $\mf=-1$ state are separated from the free expanding atoms in the $\mf=0$ state, so that the predicted spherical shell can form without perturbations from the other states. 

In order to suppress the acceleration of the outcoupled atoms due to the anti-trapping potential \eqref{eq:potential:0:spatial:dependence} during this process, the trap bottom as well as the trap frequencies are ramped down and the trap is finally switched off. For the remainder of the evolution a small magnetic-field offset is maintained to ensure a proper quantization axis for the hyperfine sublevels. The magnetic field gradient \eqref{eq:magnetic:field:gradient} serves the purpose of removing the atoms in the $\mf=-1$ state from the detection area. The timing of the magnetic gradient is chosen so that the particle densities have dropped sufficiently to minimize interactions between the different hyperfine states when passing through each other. Note also that the momentum shift for the $\mf=0$ state caused by the magnetic gradient field is negligible. 

Before turning to the results of our numerical simulations, we verify that the physical parameters that we have chosen for our model sequence are in accordance with the conditions on spatial- and state-selectivity derived in section~\ref{sec:theoretical:basics}. In addition, we also evaluate for this setup the strength of the anti-trapping potential for the $\mf = 0$ state. 

Firstly, in order to outcouple the atoms only locally, our system needs to fulfill the condition \eqref{eq:sharp:resonance:condition} as discussed in \ref{app:sharp:resonance}. When we apply the result \eqref{eq:chem:pot:numeric:result} for $\mu_{-1}/\hbar$ and the value for $\ORabi$ listed in Tab. \ref{tab:parameters:sample:sequence} to Eq. \eqref{eq:sharp:resonance:condition} divided by $(2\pi\hbar\,\mathrm{Hz})^2$ we obtain
\begin{equation}\label{eq:result:condition1}
 1.0 \cdot 10^5 \gg 8.1 \cdot 10^3 \,.
\end{equation}
Hence, the Rabi frequency that we have used should indeed be small enough to allow a spatial-selective outcoupling.

Secondly, to avoid transferring atoms from the untrapped to the anti-trapped state, the condition \eqref{eq:state:selective:condition} derived in \ref{app:state:selective} needs to hold true. By inserting the chemical potential \eqref{eq:chem:pot:numeric:result} and the value of the minimum of the magnetic field given by Tab. \ref{tab:parameters:sample:sequence} into Eq. \eqref{eq:state:selective:condition} we get
\begin{equation}\label{eq:result:condition2}
 16 \,\mathrm{G}^2 \gg 2.2 \,\mathrm{G}^2 \,.
\end{equation}
Thus, the asymmetry between the transition frequencies $\omega_{-1,0}$ and $\omega_{0,+1}$ is large enough to suppress sufficiently the unwanted outcoupling to the anti-trapped state. 

As a last check we compare the ratio of the trapping potentials for the $\mf=0$ and the $\mf=-1$ state, which can be easily derived from Eq.~\eqref{eq:potential:0:spatial:dependence} and yields
\begin{equation}\label{eq:result:ratio:potentials}
 \frac{|V_{\mathrm{trap},0}(\x,t)|}{|V_{\mathrm{trap},-1}(\x,t)|} \approx 3.3 \cdot 10^{-3} \,.
\end{equation}
Although this ratio is quite small, we will see in section \ref{sec:isotropy} that the anti-trapping potential for the $\mf=0$ state still has a relevant impact on the output of the space atom laser. In principle, this effect could be reduced by further lowering the value of the magnetic trap bottom $B_\mathrm{bot}$. However, in doing so, we would reduce the validity of the inequality~\eqref{eq:result:condition2}, which follows from Eq.~\eqref{eq:state:selective:condition}. Hence, the value of $B_\mathrm{bot}$ listed in Tab.~\ref{tab:parameters:sample:sequence} is chosen as a compromise to fulfill both conditions as good as possible. 

In summary, the parameters that we have chosen for our setup fulfill all necessary requirements for a spatially-dependent as well as state-selective outcoupling. In the next subsection we present the results of this outcoupling process.

\subsection{Time evolution}
The initial state for the time evolution was found by numerically solving the time-independent GP equation \eqref{eq:time:independent:GP} by an imaginary time propagation method \cite{Auer2001}. The dynamics of the three hyperfine sublevels $\psi_{\mf}(\x,t)$ governed by the coupled GP equations \eqref{eq:coupled:GPE:rot:wave:approx} and the model sequence described in section \ref{subsec:setup} was then computed by a Dormand-Prince embedded method \cite{Press2007}. In order to avoid reflections of the outgoing wavefunctions at the edge of the numerical grid, we made use of suitable imaginary absorption potentials \cite{Neuhauser1989}.

\begin{figure}[h]
\begin{center}
 \includegraphics[scale=0.8]{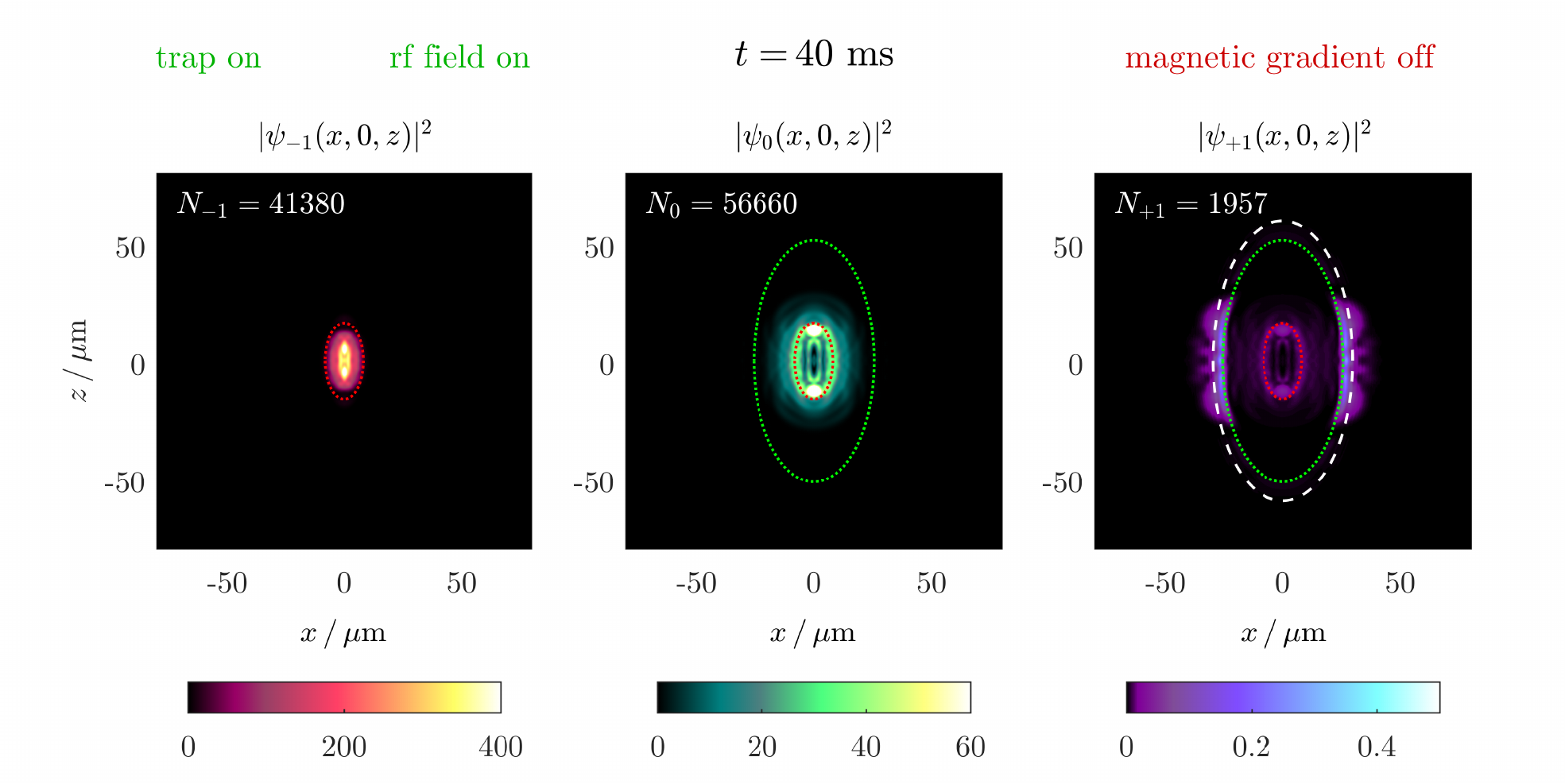}
 \caption{Density distributions of the trapped (left image), untrapped (central image) and anti-trapped state (right image) on the $xz$-plane for the time $t = 40\,\mathrm{ms}$ during the model sequence of Fig.~\ref{fig:Drf:sequence}. 
  The red dotted line displays the area where the resonance condition, Eq.~\eqref{eq:reso:condition:-1:0}, for the transition between the $\mf = -1$ and $\mf = 0$ sublevels is fulfilled, while the green line shows the resonance area for the transition between the $\mf = 0$ and $\mf = +1$ sublevels given by Eq.~\eqref{eq:reso:condition:0:+1}. 
 For numerical reasons an imaginary absorption potential was implemented for the anti-trapped state, which takes on nonzero values outside of the white dashed line, but does not influence the dynamics of the other states. 
During the time evolution an increasing number of atoms is transferred to the $\mf = 0$ sublevel and expands away from the center of the trap, while there are also losses to the $\mf = +1$ state. After the end of the outcoupling process at $t=90\,\mathrm{ms}$ the remaining atoms in the $\mf = -1$ sublevel are removed by a magnetic field gradient.
A video displaying the whole time evolution can be found in the online version of this paper. For the sake of visibility the color maps are rescaled several times during the video to compensate for the decreasing density due to the expansion of the atom cloud.}
 \label{fig:density:series:test:sequence}
\end{center}
\end{figure}

The non-integrated density distributions of all three states on the $xz$-plane are displayed in Fig.~\ref{fig:density:series:test:sequence} for the time $t = 40\,\mathrm{ms}$, which lies in the middle of the rf-outcoupling sequence displayed in Fig.~\ref{fig:Drf:sequence}. 
A movie with the full time evolution is accessible in the online version of this article.

During the outcoupling process more and more atoms are transferred from the trapped to the untrapped state which slowly expands. In this stage the wave functions of these two states resemble the slightly elongated shape of the initial state as shown in Fig.~\ref{fig:density:series:test:sequence} and undergo small periodic oscillations, which indicate the onset of collective excitations within the system. A small fraction of the BEC is also coupled to the anti-trapped state and is rapidly accelerated away from the center of the trap. 

After the rf field has been switched off at $t = 90 \,\mathrm{ms}$ the remaining atoms in the $\mf=-1$ state are accelerated along the negative $z$-axis by a magnetic field gradient at $t = 110 \,\mathrm{ms}$ and move out of the detection area. The outcoupled atoms in the $\mf=0$ state on the other hand are only marginally affected by the magnetic gradient and form a nearly spherically symmetric shell that slowly expands along all spatial directions away from the initial BEC. 

\begin{figure}[h]
\begin{center}
 \includegraphics[scale=0.9]{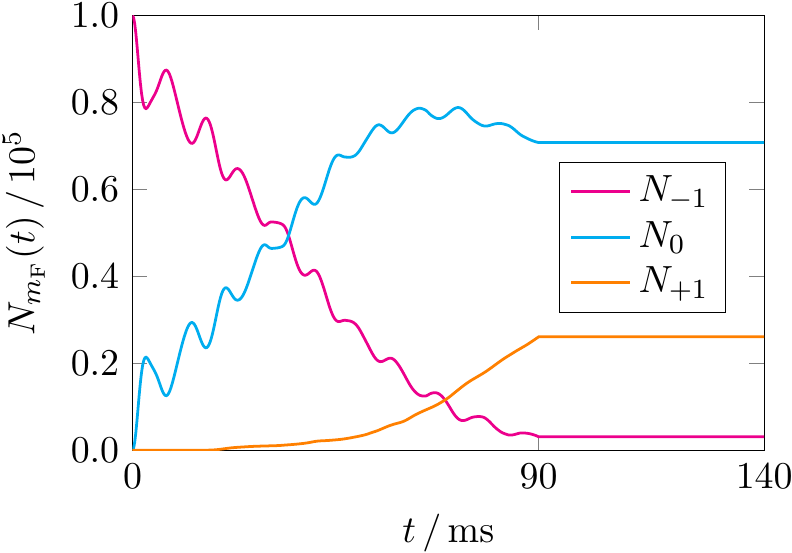}
 \caption{Change of the number of particles in each hyperfine sublevel as a function of time for the model sequence of Fig.~\ref{fig:Drf:sequence}. Initially all atoms occupy the $\mf=-1$ sublevel, which is emptied in a continuous, slightly oscillatory manner in the course of the outcoupling process. The number of particles in the $\mf=0$ sublevel grows over time, but towards the end it decreases slightly and the particle number in the $\mf=+1$ state rises. After the rf field is switched off at $t = 90 \,\mathrm{ms}$ the occupation numbers remain constant till the end of the evolution.}
 \label{fig:particle:numbers:test:sequence}
\end{center}
\end{figure}

Figure~\ref{fig:particle:numbers:test:sequence} displays the corresponding time dependence of the number of particles in each sublevel. 
Initially, all the atoms are in the $\mf=-1$ sublevel. During the outcoupling process the number of atoms in the untrapped state grows quite continuously with small oscillations modulating the particle gain, while the number of trapped atoms is reduced by the same amount. These oscillations are a consequence of the collective oscillations introduced into the system by the outcoupling process. 

Close to the end of the rf coupling the number of particles in the anti-trapped state increases since more atoms approach the resonance area between the $\mf=0$ and the $\mf=+1$ sublevel indicated by the green dotted line in Fig.~\ref{fig:density:series:test:sequence}. Consequently, the amount of particles in the $\mf=0$ state decreases slightly. 
After the rf field is switched off at $t = 90 \,\mathrm{ms}$ the number of particles stays constant in all sublevels for the rest of the evolution. 

In summary, we were able to empty the $\mf=-1$ sublevel nearly completely and to transfer $71\%$ of the atoms to the $\mf=0$ sublevel at the end of the outcoupling process with a loss of $26\%$ to the unwanted $\mf=+1$ sublevel. We emphasize that our model sequence represents a trade-off between transfer efficiency, expansion velocity and isotropy of the output state and that even higher transfer rates can be easily reached with a setup specifically chosen for that purpose.

\subsection{Analysis of the output state}
\begin{figure}[h]
\begin{center}
 \includegraphics[scale=0.9]{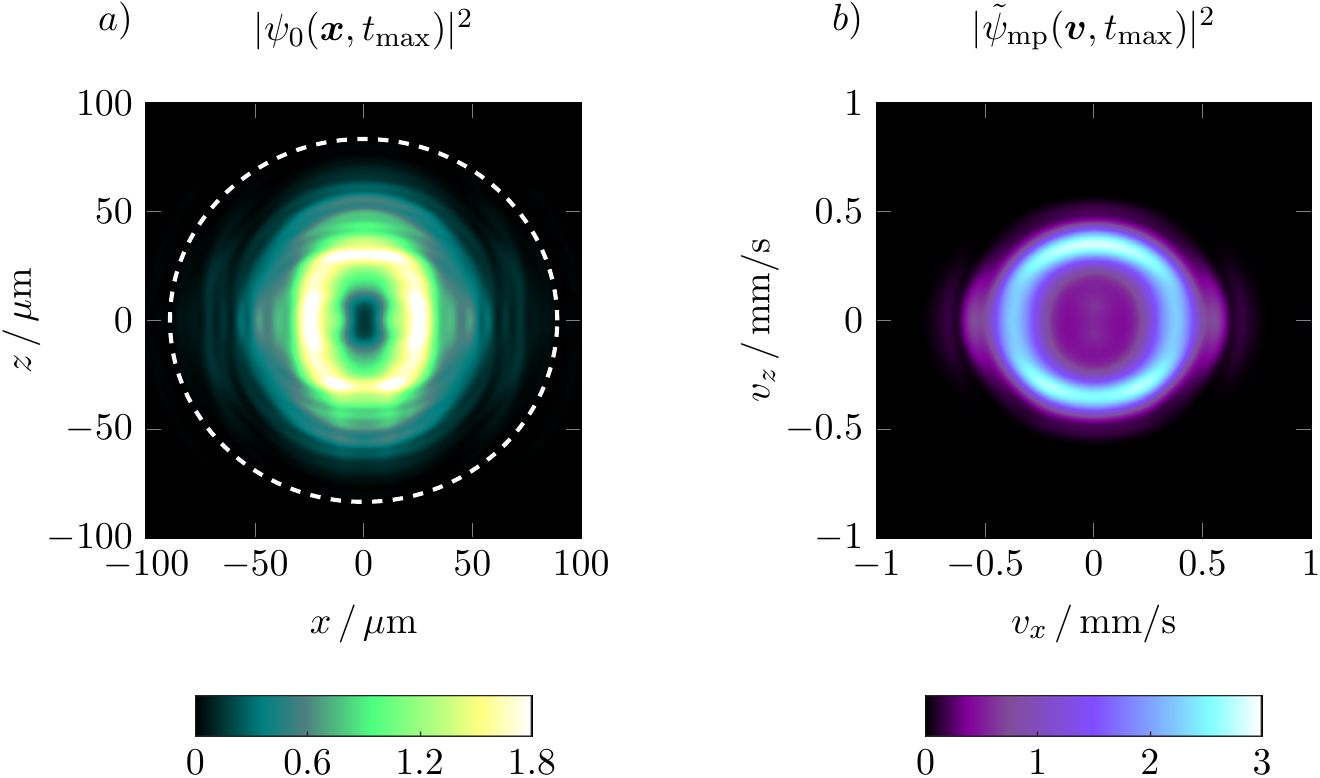}
 \caption{Final density distributions in position and momentum space of the untrapped state for the model sequence of Fig.~\ref{fig:Drf:sequence}. (a) The spatial density $|\psi_0(\x,t_{\mathrm{max}})|^2$ displayed in the $xz$-plane is rather isotropic with a shell-like main peak. The white dashed line indicates the separation ellipsoid used to distinguish between the main peak and faster expanding contributions. (b) The corresponding momentum density $|\tilde\psi_\mathrm{mp}(\vec{v},t_{\mathrm{max}})|^2$ for the main peak is also quite isotropic with a slight elongation along the $v_x$-axis.}
 \label{fig:final:space:momentum:distribution}
\end{center}
\end{figure}

The final density distributions of the untrapped state in position and momentum are depicted in Fig.~\ref{fig:final:space:momentum:distribution}. For convenience the momentum distribution is displayed as a velocity distribution making it easier to extract the expansion velocities of the space atom laser. Both the spatial and the momentum density distribution show a fairly isotropic structure, which is in good agreement with the predictions made in section \ref{subsec:output:state}, and discussed in more detail in section \ref{sec:isotropy}.

The velocity distribution $\tilde\psi_\mathrm{mp}(\vec{v},t_{\mathrm{max}})$ is calculated by means of the Fourier transform
\begin{equation}
 \tilde\psi_\mathrm{mp}(\vec{v},t_{\mathrm{max}}) =  \left(\frac{M}{2\pi\hbar}\right)^{3/2} \displaystyle\int\limits_{V_\mathrm{mp}} \Mdiff^3 x \,\Me^{- \Mi\,M\,\vec{v}\cdot \x / \hbar} \,\psi_0(\x,t_\mathrm{max}) \;,
\end{equation}
where we limit the integration to the volume $V_\mathrm{mp}$ around the main peak, which is restricted by the white dashed ellipse in Fig.~\ref{fig:final:space:momentum:distribution}a. The semi-principal axes of this ellipse, which in three dimensions is an ellipsoid, are defined by the points where the density $|\psi_0(\x,t_{\mathrm{max}})|^2$ has decreased to $10^{-3}$ of its maximum value along the respective axis. This procedure allows us to reliably compare our results with future experimental data: since in an experiment only the density around the main peak would produce a large enough signal to be measured with state-of-the-art detection schemes, the momentum distribution should also be restricted to the main peak. 
Typically the atoms outside of the main peak have a higher velocity, which can be caused either by off-resonant outcoupling in a region with higher spatial density and, hence, larger interaction energy or by outcoupling atoms that undergo collective oscillations of the system and, thus, already have a higher momentum. In summary, by restricting to the main peak, we ensure that our results are independent of these experimentally non-detectable, fast-moving atoms and thus are more useful for experimentalists.

To further quantify the output of the space atom laser we consider the number of particles within the main peak  
\begin{equation}\label{eq:particle:number:main:peak}
 N_\mathrm{mp} = \int\limits_{V_\mathrm{mp}} \Mdiff^3 x \,\left|\psi_{0}(\x,t_\mathrm{max})\right|^2
\end{equation}
and the average 3D-expansion velocity of the atoms in the main peak 
\begin{equation}\label{eq:def:average:velocity}
 \bar{v}_\mathrm{mp} = \frac{\displaystyle\int \Mdiff^3 v \,|\vec{v}| \,\left|\tilde\psi_\mathrm{mp}(\vec{v},t_{\mathrm{max}})\right|^2}{\displaystyle\int \Mdiff^3 v \,\left|\tilde\psi_\mathrm{mp}(\vec{v},t_{\mathrm{max}})\right|^2} \;.
\end{equation}
The values of these two quantities for the model sequence are listed in Tab.~\ref{tab:results:model:sequence} together with the width of the shell-like velocity distribution $\tilde\psi_\mathrm{mp}(\vec{v},t_{\mathrm{max}})$ obtained by a Gaussian fit along the $v_x$- and $v_z$-axes. The momentum overlap fidelity is discussed in the next section as a measure for the isotropy of the distribution. Table~\ref{tab:results:model:sequence} also displays the corresponding values obtained for a free expansion of the initial BEC by completely switching off the trap without using any rf field. In this case, all particles remain obviously in the $\mf=-1$ state and all the relevant quantities are, therefore, calculated for this state. 
\begin{table}[h]
\begin{center}
\caption{Comparison of the numerical results of the model sequence of Fig.~\ref{fig:Drf:sequence} for the space atom laser with the expansion of the initial BEC after a sudden release from the trap.}
\begin{small}
\begin{tabular}{lcc}
\\ \hline\hline \\
\vspace{0.3cm} &  space atom laser &  sudden trap release  \\ 
atoms in the main peak: $N_\mathrm{mp}$ &  $65 \,479$ & $99\,945$   \\
average 3D-expansion velocity: $\bar{v}_\mathrm{mp}$ &  $410\, \mu$m/s &  $864 \, \mu$m/s \\
width of the velocity shell: $(\sigma_{v_x}, \,\sigma_{v_z})$ & $(113, \,67) \,\mu\mathrm{m}/\mathrm{s}$ & -- \\
momentum overlap fidelity: $F[\tilde\psi_\mathrm{mp}]$ &  $0.961$  &  $0.842$ \\
\\ \hline\hline 
\end{tabular}
\end{small}
\label{tab:results:model:sequence}
\end{center}
\end{table}

There are of course some particle losses when using a rf-outcoupling method compared to releasing all atoms from the trap by a sudden switch-off. Nevertheless, our test sequence demonstrates a good transfer efficiency of $N_\mathrm{mp}/N = 0.65$ from the initial BEC to the main peak of the space atom laser, which confirms the state-selectivity of our method. In principle, even better outcoupling efficiencies can be reached by using a higher trap bottom $B_\mathrm{bot}$, but this strategy leads to increasing expansion velocities due to the anti-trapping potential \eqref{eq:potential:0:spatial:dependence} for the $\mf=0$ state. 

For the space atom laser we obtain a substantially lower expansion velocity $\bar{v}_\mathrm{mp}$ compared with a BEC freely expanding after a sudden release. This effect is due to the different release mechanisms used in these two scenarios: in the case of a BEC suddenly released from the trap the entire interaction energy is transformed to kinetic energy during the subsequent expansion dynamics. However, with the space atom laser we continuously outcouple atoms from the very edge of the BEC, where the interaction energy density is low. This release over time ensures that in total far less interaction energy is converted to kinetic energy and that the average expansion velocity is consequently much smaller than for a sudden release of the whole BEC. 

We emphasize that $\bar{v}_\mathrm{mp}$ represents the average expansion velocity and not the width of the velocity distribution. From Fig.~\ref{fig:final:space:momentum:distribution} one can deduce that the width of the shell-like velocity distribution is much smaller than $\bar{v}_\mathrm{mp}$.
By fitting a Gaussian to a linear cut through the velocity distribution along the $v_x$- and $v_z$-axes, respectively, we obtain the velocity widths displayed in Tab.~\ref{tab:results:model:sequence}. When we relate these velocity widths to an effective temperature
\begin{equation}\label{eq:effective:temperature}
 T_{\mathrm{eff},i} = M \sigma_{v_i}^2/k_\mathrm{B}
\end{equation}
we obtain $T_{\mathrm{eff},x} = 133\,\mathrm{pK}$ and $T_{\mathrm{eff},z} = 47\,\mathrm{pK}$, which are deep in the picokelvin regime. The small velocity spread within the velocity shell of the space atom laser is comparable to state-of-the-art experiments \cite{Muentinga2013,McDonald2013,Kovachy2015} employing delta-kick collimation to reduce the momentum width of the atomic cloud. Therefore, our rf-outcoupling method provides another independent approach of preparing ultra-cold atoms in the sub-nanokelvin regime.

Most of the analysis of the output state throughout this paper is performed in terms of the momentum distribution, which contains not only the average 3D-expansion velocity and the width of the shell-like velocity distribution, but also valuable information about the directionality and the isotropy of the space atom laser.
Note that given an isotropic distribution, how narrow the shell in momentum space can be is limited by the inverse of the radial width of the shell in position representation. This limitation corresponds to the Heisenberg uncertainty relation for position and momentum applied to a spherically symmetric wave function. In absence of gravity, and for atoms freely propagating after the outcoupling, the radial extent is determined by the radial expansion velocity $\bar{v}_\mathrm{mp}$ times the total outcoupling time $t_{\mathrm{rf}}$. Hence, taking into account that $\Delta\omega = \Delta E / \hbar = (\Delta p / \hbar) \, \bar{v}_\mathrm{mp}$, the aforementioned limit can be equivalently written in terms of a frequency linewidth and the total outcoupling time, and it is directly related to Heisenberg's uncertainty relation for time and energy. These considerations are analogous to the results obtained for the one-dimensional case in Ref.~\cite{Johnsson2007}, where various aspects connected with the linewidth of atom lasers were investigated in detail.

In principle, very narrow linewidths can be achieved for sufficiently weak outcoupling so that the outcoupling time $t_{\mathrm{rf}}$ becomes very large and nonlinear interactions in the dilute cloud of outcoupled atoms are negligible. However, this would typically lead to rather small atom fluxes that would severely limit the interest of such atom lasers and compromise absorption imaging of the emitted atomic cloud. On the other hand, for the intermediate outcoupling strengths considered here higher fluxes are possible, but additional effects such as nonlinear interactions during the outcoupling process prevent the Fourier limit for the linewidth that has been mentioned above from being reached.

\section{Isotropy of the outcoupled shell}\label{sec:isotropy}
In this section we have a closer look at the isotropy of the output state of the space atom laser. Since for a sufficiently long free evolution time the spatial density distribution is determined by the momentum distribution, we focus throughout this section on the momentum distribution. 

\subsection{Momentum overlap fidelity}
As a measure of the isotropy of a given state we introduce the \emph{momentum overlap fidelity}
\begin{equation}\label{eq:momentum:overlap:fidelity}
 F\left[\tilde\psi_\mathrm{mp}\right] = \frac{\displaystyle\int \Mdiff v_x \,\Mdiff v_z  \,\left|\tilde\psi_\mathrm{mp}(v_x,0,v_z,t_{\mathrm{max}})\right| \,\left|\tilde\psi_\mathrm{mp}(v_z,0,v_x,t_{\mathrm{max}})\right|}{\displaystyle\int \Mdiff v_x \,\Mdiff v_z \,\left|\tilde\psi_\mathrm{mp}(v_x,0,v_z,t_{\mathrm{max}})\right|^2} \leq 1 \;,
\end{equation}
which calculates the overlap between $|\tilde\psi_\mathrm{mp}|$ and the same state rotated by $90$ degrees on the $v_x v_z$-plane. For a perfectly isotropic state we have $F = 1$. The advantage of using $F[\tilde\psi_\mathrm{mp}]$ lies in the fact that this measure is sensitive to the local isotropy of the momentum distribution and does not just compare integrated quantities.

Clearly, Eq.~\eqref{eq:momentum:overlap:fidelity} only measures the four-fold symmetry of the momentum distribution and would also imply $F=1$ for a four-pointed star or a cross-shaped distribution. However, for magnetically trapped BECs outcoupled in microgravity the spatial and momentum distributions typically tend towards elliptical shapes as illustrated in Figs.~\ref{fig:final:space:momentum:distribution} and \ref{fig:momentum:distribution:comparison}. The anisotropy of such distributions can be reliably characterized by the momentum overlap fidelity presented here. 

In Tab.~\ref{tab:results:model:sequence} the momentum overlap fidelity for our model sequence is compared with the value corresponding to a BEC freely expanding after sudden release. Clearly, our rf-outcoupling method leads to a much more isotropic momentum distribution. This isotropy is one of the unique properties of the space atom laser compared to other sources of ultra-cold atoms, especially when taking into account that for the space atom laser it emerges naturally during the outcoupling process, as discussed in section \ref{subsec:output:state}.

\subsection{Dependence on the Rabi frequency}
\begin{figure}[h]
\begin{center}
 \includegraphics[scale=0.9]{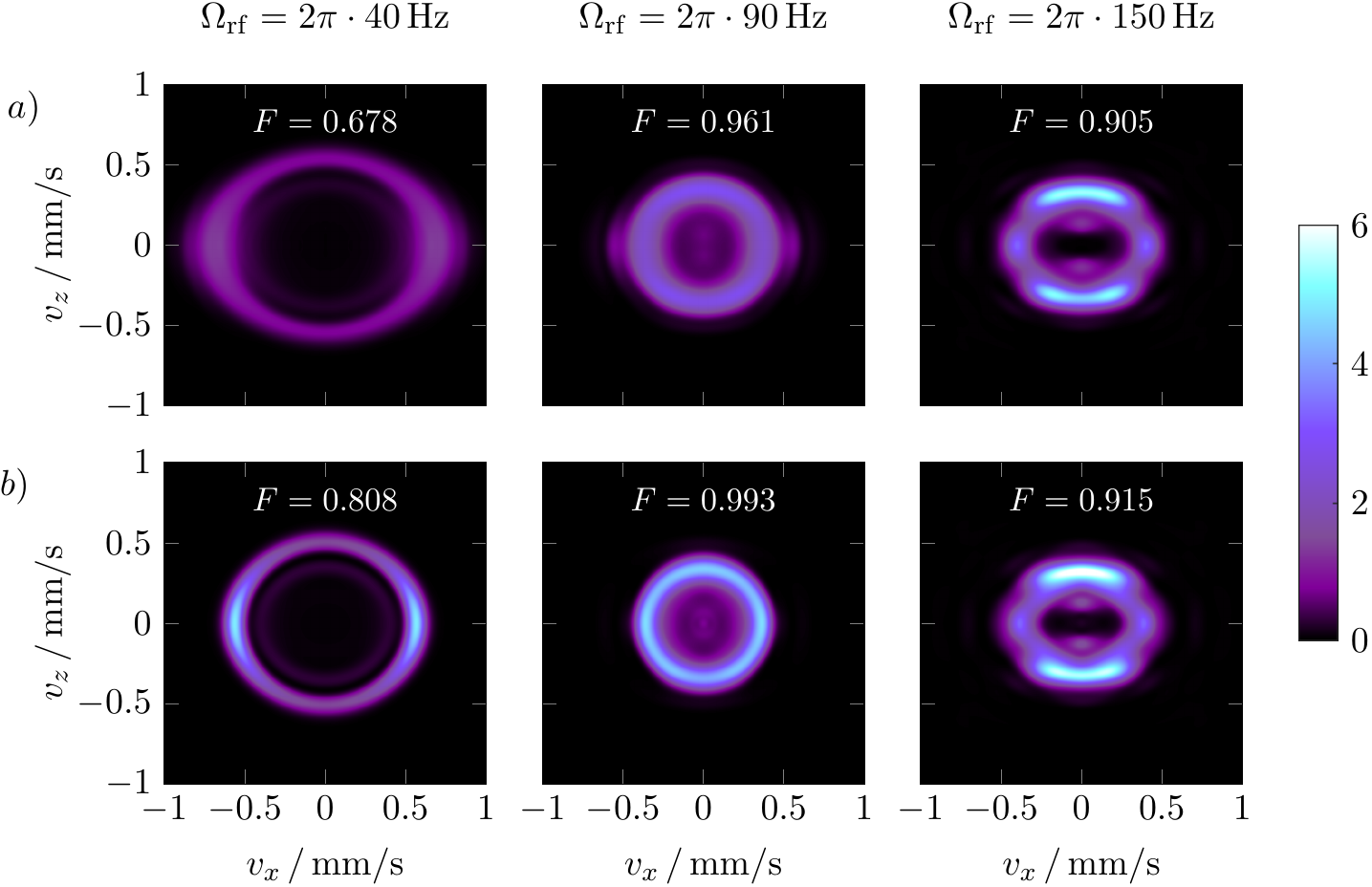}
 \caption{Momentum density distributions $|\tilde\psi_\mathrm{mp}(\vec{v},t_{\mathrm{max}})|^2$ for the main peak of the space atom laser together with the momentum overlap fidelity $F[\tilde\psi_\mathrm{mp}]$ for different Rabi frequencies whilst (a) taking into account the anti-trapping potential, Eq. \eqref{eq:potential:0:spatial:dependence}, for the $\mf=0$ state and (b) setting $V_{\mathrm{trap},0}(\x,t) = 0$. For $\ORabi = 2\pi\cdot 90\,\mathrm{Hz}$ the momentum distribution possesses a nearly isotropic, ring-like shape. For smaller Rabi frequencies the ring gets wider and the momentum density is distributed anisotropically across the ring, while for higher Rabi frequencies the momentum distribution gets strongly deformed. For small and medium Rabi frequencies the anti-trapping potential for the $\mf=0$ state (a) leads to a broadening of the momentum distribution, whereas for $\ORabi = 2\pi\cdot 150\,\mathrm{Hz}$ this effect is only minor. }
 \label{fig:momentum:distribution:comparison}
\end{center}
\end{figure}

As already mentioned earlier, several experimental parameters do have an impact on the isotropy of the outcoupled state, in particular the Rabi frequency and the strength of the anti-trapping potential for the $\mf=0$ state. In Fig.~\ref{fig:momentum:distribution:comparison} the momentum distributions and the corresponding values of $F[\tilde\psi_\mathrm{mp}]$ are displayed for three different Rabi frequencies. In Fig.~\ref{fig:momentum:distribution:comparison}a we have used exactly the same sequence as described in section~\ref{sec:sample:test:sequence}, while the results displayed in Fig.~\ref{fig:momentum:distribution:comparison}b were obtained by setting $V_{\mathrm{trap},0}(\x,t) = 0$ during the outcoupling process. 

As can be easily seen, the choice of the correct Rabi frequency is crucial for generating an isotropic output state. Even though Fig.~\ref{fig:momentum:distribution:comparison} shows in principle a ring-like shape for all three Rabi frequencies, the density distribution across this ring depends strongly on $\ORabi$. For small Rabi frequencies ($\ORabi = 2\pi\cdot 40\,\mathrm{Hz}$) the momentum density along the $v_x$-axis is the highest, while for large Rabi frequencies ($\ORabi = 2\pi\cdot 150\,\mathrm{Hz}$) most atoms have a momentum pointing along the $v_z$-axis. For the intermediate value $\ORabi = 2\pi\cdot 90\,\mathrm{Hz}$ the momentum density is distributed equally on all spatial directions forming a truly isotropic distribution. 

This phenomenon can be explained in the following way: for small Rabi frequencies the outcoupling is weak enough not to excite collective modes of the trapped BEC. Instead, the atoms are outcoupled at the edge of the BEC, gain a specific amount of kinetic energy equal to the interaction energy \eqref{eq:interaction:potential:-1:resonance} and propagate orthogonally away from the surface of the trapped BEC. Since this surface is an ellipsoid elongated along the $z$-axis as displayed in the left image of Fig.~\ref{fig:density:series:test:sequence}, most atoms have a larger velocity along the $x$- and $y$-axes than along the $z$-axis and this leads to the momentum distribution displayed in the first image of Fig.~\ref{fig:momentum:distribution:comparison}. 

As the Rabi frequency is increased the resonance for transitions from the $\mf=-1$ to the $\mf=0$ hyperfine state gets wider, so that more atoms are outcoupled at once. This relation leads to the onset of collective oscillations of the trapped BEC and also changes the distribution of directions in which the outcoupled atoms propagate away from the condensate. Since $\omega_z$ is the smallest trap frequency the spatial density of the trapped BEC has the lowest slope along the $z$-axis and thus an increase in the resonance width directly leads to an overproportional gain of atoms being outcoupled along the $z$-axis compared with the $x$- and $y$-axes. As a consequence, by increasing the Rabi frequency the output of the space atom laser changes from a regime where most atoms propagate along the tight-confining axes to a regime, where a majority of the atoms are outcoupled along the weakly-confining axis, as displayed in Fig.~\ref{fig:momentum:distribution:comparison}. Obviously, for intermediate Rabi frequencies this effect results in a more isotropic momentum distribution.

However, as our numerical simulations have shown, the true dynamics of the space atom laser is more complex due to the influence of collective oscillations, which distort the shape of the trapped BEC and thus lead to substantial deviations from the ellipsoidal shape. Consequently, for large Rabi frequencies the resulting momentum distributions lose their ring-like shape as displayed on the right column of Fig.~\ref{fig:momentum:distribution:comparison}. Nevertheless, the above considerations are valuable to explain the general dependence of the shape of the momentum distribution on the Rabi frequency.

\subsection{Influence of the anti-trapping potential for the $\mf=0$ state}
In addition, as shown in Fig.~\ref{fig:momentum:distribution:comparison}, the strength of the anti-trapping potential for the $\mf=0$ state, given by Eq. \eqref{eq:potential:0:spatial:dependence}, also has a strong influence on the isotropy of the output state, especially for small and intermediate Rabi frequencies. Indeed, when we compare the results for the correct anti-trapping potential in Fig~\ref{fig:momentum:distribution:comparison}a with the ones for $V_{\mathrm{trap},0}(\x,t) = 0$ shown in Fig.~\ref{fig:momentum:distribution:comparison}b, we observe a clear broadening of the momentum distribution due to the additional acceleration away from the initial BEC. Since this acceleration is proportional to the strength of the magnetic trap, its effect is stronger along the tightly-confining axes, which leads to an anisotropic elongation of the momentum distribution along these axes and, consequently, to a reduced overlap fidelity $F[\tilde\psi_\mathrm{mp}]$. For large Rabi frequencies the atoms are coupled back and forth between the $\mf=-1$ and the $\mf=0$ states multiple times so that in this case they are not continuously exposed to the anti-trapping potential for the $\mf=0$ state and this causes substantially less broadening of the momentum distribution, as shown in the last column of Fig.~\ref{fig:momentum:distribution:comparison}. 

In summary, the anti-trapping potential for the $\mf=0$ state prevents us from reaching even higher isotropies with our rf outcoupling scheme.  
Therefore, it is advisable to reduce the impact of this potential as much as possible. According to Eq.~\eqref{eq:potential:0:spatial:dependence} lowering the trap bottom $B_\mathrm{bot}$ or the trap frequencies of the magnetic trap would help in reaching this goal. However, by lowering $B_\mathrm{bot}$, the asymmetry \eqref{eq:transition:frequencies:not:equal} between the different hyperfine states would also be reduced and this would lead to a larger loss rate to the $\mf=+1$ state during the outcoupling process. As a consequence, one needs to make a trade-off between the outcoupling efficiency and the expansion rate of the atoms. The parameters used for our model sequence constitute such a compromise and show that it is indeed possible to achieve a good performance of the space atom laser with rf outcoupling.

One possible solution to circumvent this trade-off would be to use a highly stable microwave source to drive an effectively two-level system between the $F=1,\mf=-1$ and the $F=2,\mf=0$ states as demonstrated for an earth-bound laboratory in Ref.~\cite{Oettl2005,Oettl2006}. In this case one could lower the offset magnetic field further without losing atoms to an unwanted state. As a result, the outcoupling efficiency could be maximized, while at the same time the potential due to the general Zeeman effect, which for the $F=2,\mf=0$ state is a trapping potential \cite{BergmannSchaefer1997}, would have just a minor effect on the expansion dynamics.
However, the absolute requirements on the stability and the tuneability of the microwave frequency would be the same as in the rf case, but with an offset frequency of the order $10^3$ times larger. 
Nevertheless, using microwave outcouplers in a dedicated experimental setup could be one way to suppress the higher-order Zeeman potential for the untrapped state and allow even higher isotropies. 

Another alternative for reducing the acceleration that acts on the outcoupled atoms is to superimpose an optical dipole trap which compensates the magnetic potential for the $\mf=0$ state as demonstrated in Refs.~\cite{Guerin2006,Billy2007} with a horizontal wave guide. However, transferring this concept to microgravity would require three optical beams in order to compensate the elongated magnetic trap along all spatial directions and is, therefore, quite challenging to implement.  

We conclude this section by emphasizing that the effect of a constant output velocity across the resonance area
was already mentioned in Ref.~\cite{Naraschewski1997} in the context of a theoretical study of an atom laser with a continuous repumping mechanism that allowed to keep the wave function of the trapped state constant during the dynamics. 
Within this paper we have shown that an isotropic and narrow velocity distribution can be realized even when the trapped state is continuously emptied over time and undergoes small collective oscillations during the outcoupling process. 
In addition, we have also discussed in detail the origin of possible deviations from this isotropic shape depending on the Rabi frequency. 
Finally, we stress that despite the additional acceleration due to the general Zeeman effect our rf-outcoupling scheme for the space atom laser allows to convert an initially trapped, anisotropic BEC into a nearly isotropic, freely expanding one.

\section{Influence of varying parameters}\label{sec:parameter:scans}
The results presented in sections \ref{sec:sample:test:sequence} and \ref{sec:isotropy} demonstrate the properties of the space atom laser for an outcoupling sequence that was optimized for a particular set of system parameters. For different setups this procedure can be repeated and a suitable outcoupling scheme can be found for each case. However, in practice not all parameters such as the number of particles, or the value of the magnetic-field offset are known at arbitrary precision. In fact, they might vary substantially from shot to shot, or over a longer period. It is, therefore, of great importance to understand the behavior of the system under different conditions. 

In this section we study the influence of such varying parameters on the outcoupling results. At first we change the initial number of particles and the Rabi frequency associated with the rf field, but apply otherwise exactly the same outcoupling sequence shown in Fig.~\ref{fig:Drf:sequence}. In a second step we also vary the magnetic-field offset. To measure the effect of these changes on the outcoupling process, we look at the outcoupled fraction $N_\mathrm{mp}/N$ for the main peak of the space atom laser, the corresponding average expansion velocity $\bar{v}_\mathrm{mp}$ and the momentum overlap fidelity $F[\tilde\psi_\mathrm{mp}]$.
We conclude this section with a brief discussion of the position dependence of the Rabi frequency across the outcoupling region due to spatial variations of the rf-field strength and the magnetic-field alignment.

\subsection{Varying number of particles and Rabi frequency}\label{subsec:vary:num:and:Rabi}
\begin{figure}[h]
\begin{center}
 \includegraphics[scale=0.9]{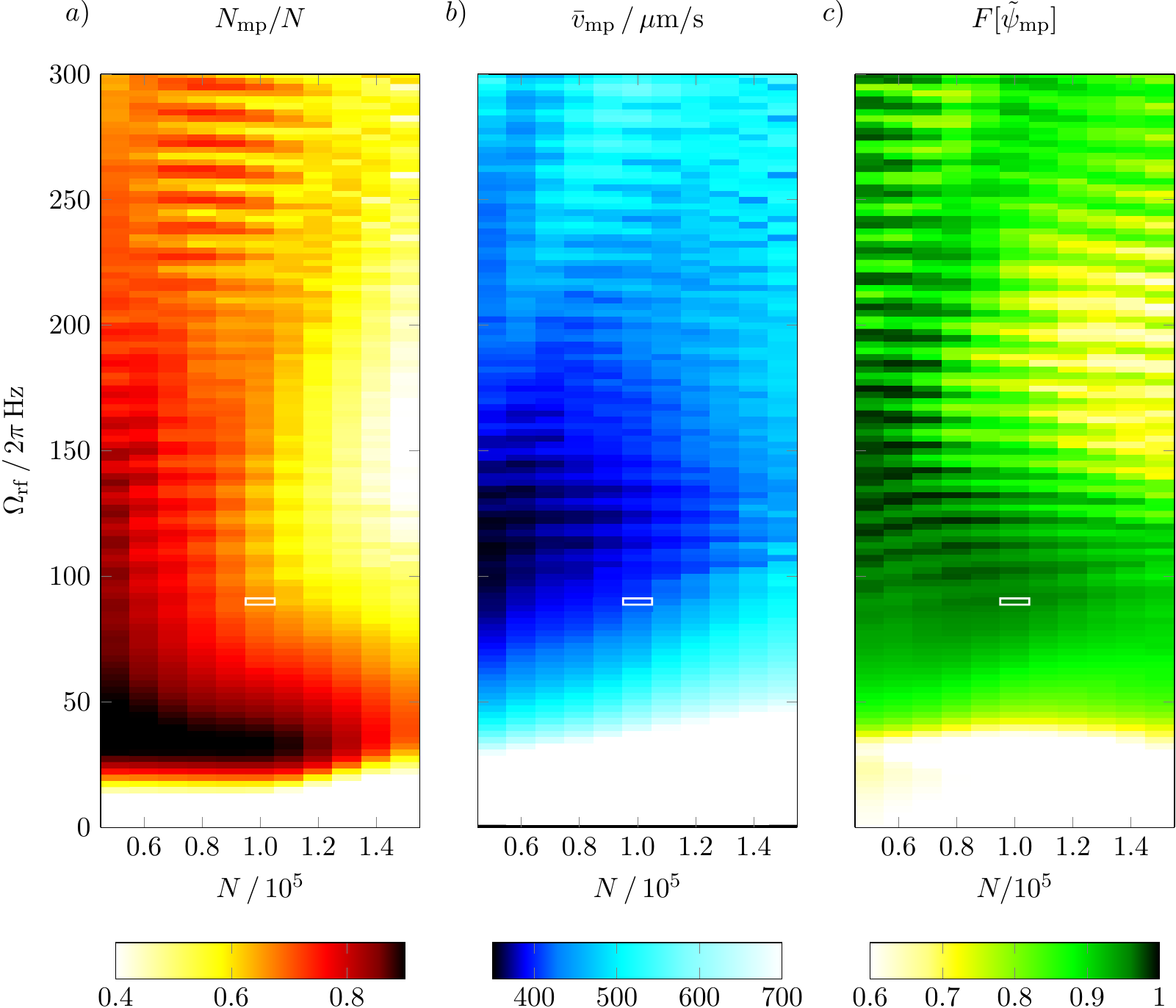}
 \caption{Outcoupled fraction (a), average expansion velocity (b) and momentum overlap fidelity (c) for varying numbers of particles in the initial BEC and different values of the Rabi frequency. Dark red colors indicate a high outcoupling efficiency (a), dark blue colors correspond to low expansion velocities (b) and dark green colors to a high isotropy (c) of the output state. In all three plots the white rectangle highlights the parameters used in the model sequence presented in section \ref{sec:sample:test:sequence}. For large values of $\ORabi$ all three measures show an oscillatory behavior. The outcoupled fraction is maximal for small $N$ and small $\ORabi$, while it approaches zero for even smaller values of $\ORabi$ and has another minimum for large $N$ and intermediate values of $\ORabi$. The average expansion velocity gets minimal for small $N$ and intermediate values of $\ORabi$. The momentum overlap fidelity possesses a strong oscillatory behavior, but is relatively close to one and constant for intermediate values of $\ORabi$.}
 \label{fig:full_parameter_scans}
\end{center}
\end{figure}

We have performed more than one thousand simulations to study the influence of a varying number of particles in the initial BEC and different but constant coupling strengths during the outcoupling sequence. The number of particles has been varied by up to $\pm 50\%$ around $N=10^5$ with a step size of $10\%$, while the Rabi frequency has been scanned from $\ORabi = 0$ to $\ORabi = 2\pi \cdot 300 \,\mathrm{Hz}$ with a step size of $2\pi \cdot 2.5\,\mathrm{Hz}$. The outcoupling efficiency, the average expansion velocity, and the momentum overlap fidelity for all these combinations are shown in Fig.~\ref{fig:full_parameter_scans}.

To understand the rich structure of these parameter plots, we divide them into several regimes, which can be summarized as follows: (i) there is a general dependency of the outcoupling results on the initial particle number $N$, (ii) in the regime $\ORabi < 2\pi\cdot 50\,\mathrm{Hz}$ the overall performance is quite poor and (iii) for $\ORabi > 2\pi\cdot 100\,\mathrm{Hz}$ the output results oscillate with increasing Rabi frequency. The choice of parameters for the model sequence discussed in the previous sections is highlighted in Fig.~\ref{fig:full_parameter_scans} with a white rectangle and lies in a small sweet spot (iv) between these different regimes that yields good results for all three measures. 

\paragraph{(i) Impact of the particle number}
For a wide range of Rabi frequencies the results of the outcoupling process do depend on the initial particle number $N$ as displayed in Fig.~\ref{fig:full_parameter_scans}. The reason for this effect is that by varying the initial particle number $N$ the chemical potential changes and consequently the edge of the BEC is shifted with respect to the ellipsoid where the outcoupling transition is resonant. Hence, for a lower particle number the rf field is initially resonant outside of the BEC, while for a larger number of particles the resonance lies initially within the BEC. In the latter case the outcoupled atoms gain more momentum due to the higher interaction energy density inside the BEC, which explains the higher expansion velocities $\bar{v}_\mathrm{mp}$ observed for larger initial BECs (see Fig.~\ref{fig:full_parameter_scans}b). Faster moving atoms also reach the resonance area between the $\mf=0$ and the $\mf=+1$ state earlier and are, therefore, more likely to be coupled to the anti-trapped state. This loss mechanism reduces the outcoupling efficiency $N_\mathrm{mp}/N$ with growing initial particle number (see Fig.~\ref{fig:full_parameter_scans}a). 

In addition, the resonance area for the rf transition has the same shape as the magnetic trap, which is elongated along the $z$-axis. Thus, the losses to the $\mf=+1$ state are distributed anisotropicaly leading to a smaller momentum overlap fidelity for the outcoupled state as displayed in Fig.~\ref{fig:full_parameter_scans}c.

\paragraph{(ii) Shut-down for small coupling strength}
In the regime of small Rabi frequencies ($\ORabi < 2\pi\cdot 50\,\mathrm{Hz}$) the outcoupling efficiency decreases from its maximum value to zero for vanishing coupling strength. In addition, the expansion velocity is very large and the momentum overlap fidelity is minimal. 

The decreasing outcoupling efficiency can be traced back to the fact that for small Rabi frequencies the coupling between the hyperfine sublevels is too weak to fully empty the BEC from the edge to the center and, thus, the amount of outcoupled atoms approaches zero as $\ORabi$ decreases (see Fig.~\ref{fig:full_parameter_scans}a). As a consequence, the outcoupling process is slower than the ramp-down of the rf frequency, as given by Eq. \eqref{eq:Delta:rf:test:sequence}, and the resonance is thus shifted closer to the center of the BEC, where the interaction energies are larger and the outcoupled atoms gain more momentum. 

This effect is even amplified for larger BECs, since in this case the outcoupling process is resonant within the BEC already from the start. As a consequence, we observe an increase of $\bar{v}_\mathrm{mp}$ for growing $N$ and decreasing $\ORabi$ as shown in Fig.~\ref{fig:full_parameter_scans}b. The fact that for small values of $\ORabi$ the momentum overlap fidelity decreases has been discussed in section~\ref{sec:isotropy} in connection with Fig.~\ref{fig:momentum:distribution:comparison} and can be attributed to the elongated shape of the trapped BEC, which favors an anisotropic distribution of the velocity density for small Rabi frequencies. Consequently, $F[\tilde\psi_\mathrm{mp}]$ gets minimal for decreasing $\ORabi$ as displayed in Fig.~\ref{fig:full_parameter_scans}c.

\paragraph{(iii) Rabi oscillations for large coupling strength}
Another distinct feature is the oscillatory behavior observed for all three measures, which sets in for values of $\ORabi > 2\pi\cdot 100\,\mathrm{Hz}$ with a constant periodicity of approximately $10\,\mathrm{Hz}$. As shown in Fig.~\ref{fig:full_parameter_scans}, these stripe patterns are mutually correlated so that a high outcoupling efficiency corresponds to large expansion rates and low isotropy, while a small outcoupling efficiency is connected to smaller expansion velocities and a higher momentum overlap fidelity. 
This relation is a consequence of the increasing Rabi frequency, which allows to simultaneously transfer an ever growing fraction of the trapped state to the untrapped state and back till eventually the whole BEC undergoes full Rabi oscillations between the trapped and the untrapped state. Finally, the validity of the sharp-resonance condition \eqref{eq:sharp:resonance:condition} breaks down for large values of $\ORabi$.
In this case the outcoupling results crucially depend on the phase of the Rabi cycle transferring the atoms back and forth between the $\mf=-1$ and the $\mf=0$ state. If the rf field is switched off when a lot of atoms are in the untrapped state, then all these atoms expand together and their interaction energy gets converted into kinetic energy leading to fairly high expansion velocities. On the other hand, if only few atoms are in the untrapped state after switch-off, their interaction energy is much smaller and they consequently expand more slowly. 

In addition, with a growing Rabi frequency more collective modes are excited and deform the trapped BEC at an increasing rate. This effect leads to a less isotropic momentum distribution when most atoms are outcoupled simultaneously at the end of the sequence instead of being outcoupled over time and averaging out the collective oscillations of the BEC to a certain degree. 
Finally, when a lot of atoms are continuously transferred back and forth between the $\mf=-1$ and the $\mf=0$ state during the rf ramp-down, then far fewer atoms are coupled to the $\mf=+1$ state and thus the outcoupling efficiency as well as the isotropy of the output state increases slightly for large values of $\ORabi$, as displayed in Fig.~\ref{fig:full_parameter_scans}a and Fig.~\ref{fig:full_parameter_scans}c, respectively. 

\begin{figure}[h]
\begin{center}
 \begin{tabular}{ccc}
  \includegraphics[scale=0.9]{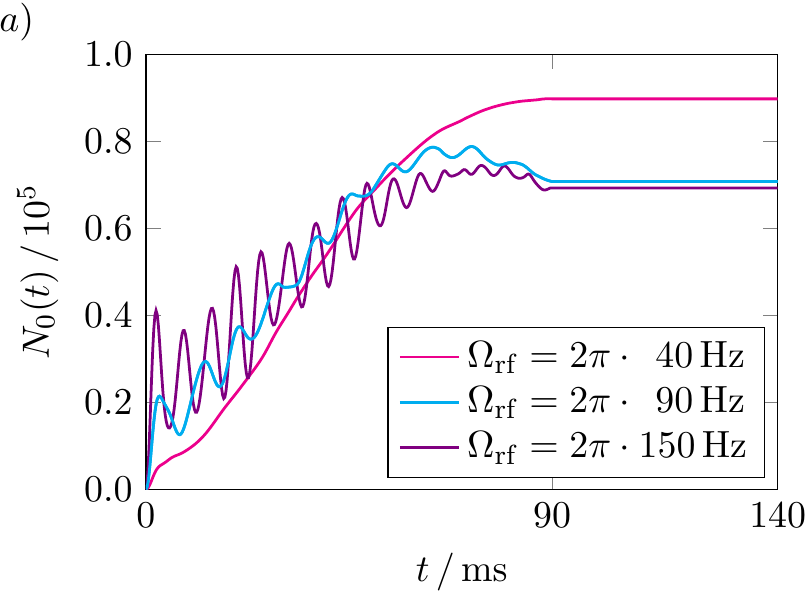} && 
  \includegraphics[scale=0.9]{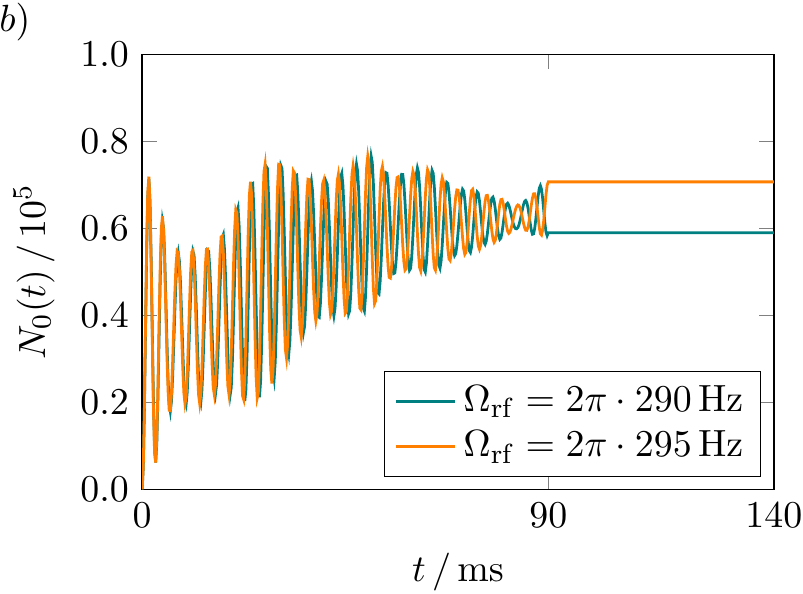} \\
 \end{tabular}
 \caption{Change over time of the number of particles in the $\mf=0$ state for different values of the Rabi frequency and a constant number of particles $N = 10^5$ for the initial BEC. For small and intermediate Rabi frequencies (a) the time dependence changes from a monotonic growth ($\ORabi = 2\pi \cdot 40\,\mathrm{Hz}$) to a gain modulated by increasing oscillations ($\ORabi = 2\pi \cdot 150\,\mathrm{Hz}$). For large Rabi frequencies (b) $N_0(t)$ is dominated by strong oscillations and its final value is very sensitive to $\ORabi$. }
 \label{fig:particle:numbers:comparison}
\end{center}
\end{figure}

To better illustrate the Rabi oscillations mentioned above, Fig.~\ref{fig:particle:numbers:comparison} shows the time dependence of the number of particles in the $\mf=0$ state for five different values of the Rabi frequency. For small Rabi frequencies ($\ORabi = 2\pi\cdot 40 \,\mathrm{Hz}$) the population of the untrapped state grows monotonically and reaches very high values.
For intermediate Rabi frequencies ($\ORabi = 2\pi\cdot 90 \,\mathrm{Hz}$ and $150\,\mathrm{Hz}$) this gain of particles is modulated by oscillations which grow in amplitude for larger values of $\ORabi$. In addition, losses to the $\mf=+1$ state lead to a slight reduction of $N_0(t)$ towards the end of the outcoupling process. 
Nevertheless, in this intermediate regime the Rabi frequency affects the outcoupling efficiency only slightly and the final value of the number of outcoupled atoms is mainly determined by the slope of the rf ramp.

This situation changes for larger values of $\ORabi$, as shown in Fig.~\ref{fig:particle:numbers:comparison}b, where roughly half of the BEC undergoes Rabi oscillations with losses, in the sense that not all atoms transferred to the untrapped state are coupled back to the trapped one. Thus, over time the average number of particles in the untrapped state increases, but with a much smaller rate compared to the intermediate regime, Fig.~\ref{fig:particle:numbers:comparison}a. In this case the final number of outcoupled atoms crucially depends on whether the rf field is switched off close to a maximum or a minimum of these Rabi oscillations. Since for our model sequence the rf field is switched on for $90\,\mathrm{ms}$ a shift of $2\pi\cdot 11\,\mathrm{Hz}$ of the Rabi frequency enables an additional full Rabi cycle within the outcoupling time, which is in excellent agreement with the stripe pattern shown in Fig.~\ref{fig:full_parameter_scans}. Hence, the difference of $2\pi\cdot 5\,\mathrm{Hz}$ of the Rabi frequency displayed in Fig.~\ref{fig:particle:numbers:comparison}b roughly corresponds to half a Rabi cycle resulting in a large difference of the final atom number $N_0(t_\mathrm{max})$.

The aforementioned Rabi oscillations between the trapped and the untrapped states can also be understood from a different but equivalent perspective, namely in the context of adiabatic dressed potentials \cite{Zobay2004,Perrin2017,Colombe2004,Lesanovsky2007}, which lead to a deformation of the external potentials for all hyperfine sublevels in the case of growing coupling strength $\ORabi$. As a consequence, for large Rabi frequencies nearly all atoms are trapped within the dressed potentials mixing the hyperfine sublevels so that there is no continuous outcoupling as long as the rf field is switched on. After swiching off the rf field, the atoms are projected back to the undressed $\mf$-states and a fraction of them expands freely. This effect sets clear limitations on the flux of continuous atom lasers and has been studied for gravity driven \cite{Robins2005,Debs2010} as well as optically guided atom lasers \cite{Bernard2011}.

\paragraph{(iv) Sweet spot for intermediate Rabi frequencies}
For intermediate Rabi frequencies ($2\pi \cdot 70\,\mathrm{Hz} < \ORabi < 2\pi \cdot 120\,\mathrm{Hz}$) the outcoupling results are overall quite good with a reasonably high outcoupling efficiency, small expansion velocities and a very isotropic momentum distribution of the output state, as displayed in Fig.~\ref{fig:full_parameter_scans}. Close enough to this sweet spot the rf coupling is strong enough to outcouple the whole BEC continuously from the edge to its center, but at the same time relation \eqref{eq:sharp:resonance:condition} is still fulfilled, so that the resonance is sharp enough to outcouple just the atoms with small interaction energies at the edge of the trapped state.
This intermediate regime is the optimal choice for operating the space atom laser since small changes of the parameters do not affect the outcoupling results much. This is especially true for the momentum overlap fidelity, which shows nearly no dependence on the initial particle number for this range of parameters. Hence, in order to observe an output state which is as isotropic as possible, choosing the experimental parameters in this intermediate regime is particularly favorable.

\subsection{Fluctuations of the offset magnetic field}
In addition to fluctuations of the number of particles, and different choices of the Rabi frequency there is another quantity which is of great importance in order to perform a space-atom-laser experiment successfully: the stability of the offset magnetic field. Magnetic fields generated by currents running through wires show small drifts and fluctuations over the course of an experiment. For the space atom laser the offset magnetic field determines the trap bottom and consequently the transition frequencies between the hyperfine sublevels. If this quantity deviates unexpectedly from its predicted value, then the applied rf frequency will not match the transition frequency anymore and no atoms will be transferred to the untrapped state. 

\begin{figure}[h]
\begin{center}
 \includegraphics[scale=0.75]{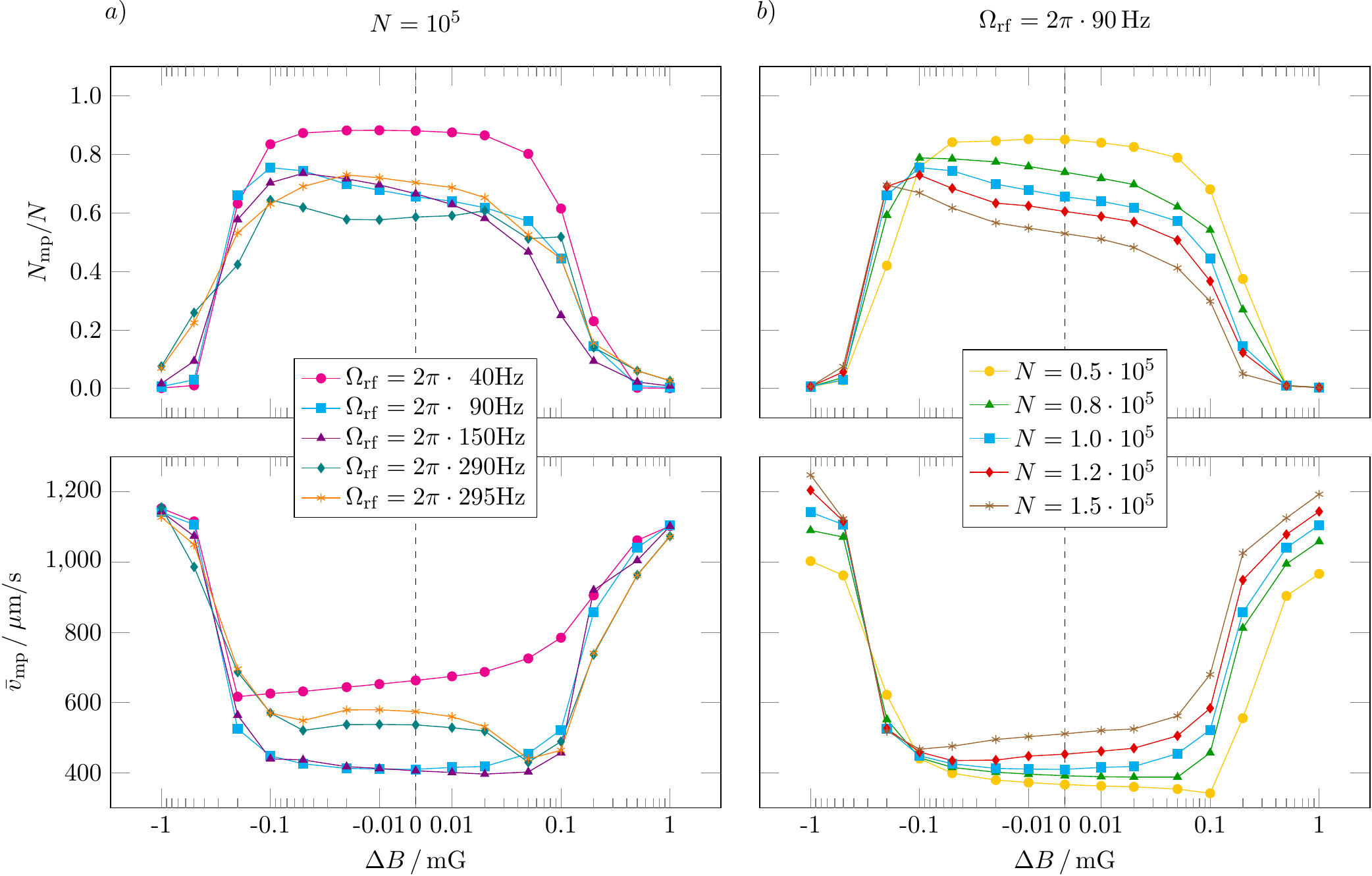}
 \caption{Outcoupled fraction (first row) and average expansion velocity (second row) as functions of a varying magnetic field for (a) different values of the Rabi frequency with the number of particles $N = 10^5$ and (b) different initial particle numbers with the Rabi frequency $\ORabi = 2\pi \cdot 90\,\mathrm{Hz}$. The scale of the magnetic field deviation $\Delta B$ is logarithmic with the range $\pm 0.01\,\mathrm{mG}$ to $\pm 1\,\mathrm{mG}$ centered around $\Delta B = 0$. For values of $|\Delta B| \leq 0.1\,\mathrm{mG}$ the outcoupled fraction is quite large and the expansion velocity is small, while for larger values of $|\Delta B|$ the outcoupling efficiency approaches zero rapidly and the expansion velocity increases.}
 \label{fig:B:offset:change_all:plots}
\end{center}
\end{figure}

To study this effect, we have performed additional simulations, where we have introduced a deviation of the magnetic field $\Delta B$, which takes on values between $-1\,\mathrm{mG}$ and $1\,\mathrm{mG}$, but has been kept constant during a single run. The remaining parameters are the same as in the previous simulations, including the rf ramp shown in Fig.~\ref{fig:Drf:sequence}. The results for different combinations of the initial particle number and the Rabi frequency are displayed in Fig.~\ref{fig:B:offset:change_all:plots}.

Figure~\ref{fig:B:offset:change_all:plots}a shows that, independently of the Rabi frequency, the output of the space atom laser is quite stable as long as the deviations of the magnetic field stay below $\pm 0.1\,\mathrm{mG}$. For larger deviations the outcoupling efficiency goes down and rapidly approaches zero, while the expansion velocity increases. This behavior is not surprising since the chemical potential of the initial BEC measured in magnetic-field strength is given by $2 \mu_{-1}/\mu_\mathrm{B} = 0.45\,\mathrm{mG}$ according to Eqs. \eqref{eq:lin:Zeeman:effect} and \eqref{eq:chem:pot:numeric:result}. Thus, deviations of the magnetic field of this order shift the resonance position of the rf transition faraway enough from where the atoms actually are to suppress any outcoupling. Hence, in order to reliably outcouple a space-atom-laser shell the deviations of the magnetic fields that generate the trap should be well below the chemical potential of the initial BEC.

For different initial particle numbers, Fig.~\ref{fig:B:offset:change_all:plots}b, we obtain in principle a similar behavior. However, here we also observe a shift along the magnetic-field deviation namely that small initial BECs yield better results (high outcoupling efficiency and low expansion rate) for positive values of $\Delta B$ than large BECs, while for negative values of $\Delta B$ the behavior is the opposite. This phenomenon can again be traced back to the chemical potential of the system, which grows with an increasing number of particles and, thus, influences the resonance position of the rf transition. 
In fact, if there are initially less particles in the trapped state, then the resonance position $\x^{\mathrm{reso}}_{-1,0}(t)$, which is determined by Eq. \eqref{eq:reso:Drf} and the rf ramp displayed in Fig.~\ref{fig:Drf:sequence}, lies outside the BEC and less atoms are hence transferred to the untrapped state. However, this gap can be closed by either reducing the rf frequency $\wrf(t)$ or by raising the offset magnetic field, which results in an increase of the offset frequency $\omega_0$. In contrast, for a large initial BEC the resonance starts within the BEC and an increase of $\wrf(t)$ or a decrease of the magnetic field shifts this resonance back to the edge of the BEC. 
Therefore, for a given chemical potential deviations of the magnetic field can be compensated by tuning the rf frequency appropriately. This dependence can also be used to systematically check for magnetic field deviations by scanning the rf frequency stepwise.

Finally, it should be noted that the results shown in Fig.~\ref{fig:B:offset:change_all:plots}b can be alternatively interpreted as providing information on the influence of a shift of the rf-field detuning $\Delta_\mathrm{rf}(t)$ instead of a change of the magnetic-field offset on the outcoupling process. Accordingly, from Fig.~\ref{fig:B:offset:change_all:plots}b we can conclude that in general the best outcoupling results can be obtained if the rf ramp starts with a frequency fulfilling the resonance condition close to the edge of the initial BEC.

\subsection{Spatial dependence of the Rabi frequency}
In addition to the effects discussed earlier in this section we have also studied the influence of a realistic, spatially dependent rf field, and of the alignment between the rf field and the magnetic moment of the trapped atoms, which both affect the value of the Rabi frequency. These calculations are based on a chip model that allows to compute the magnetic fields generated by the different wires of the CAL chip together with a ring coil for the rf field. 

Due to the relatively small size of the trapped BEC the rf field strength $B_\mathrm{rf}$ and consequently the Rabi frequency, Eq. \eqref{eq:Omega:Rabi}, undergo a relative change of less than $10^{-3}$ across the whole BEC. Hence, within the intermediate regime discussed in section \ref{subsec:vary:num:and:Rabi} this small change of the Rabi frequency has a negligible effect on the outcoupling dynamics and we can safely assume a homogeneous rf field in this case.  

When we additionally take into account the spatial dependence of the alignment of the rf field and the magnetic field, which acts as a quantization axis for the magnetic moment of the atoms, we obtain a richer dependence of the Rabi frequency on position. However, the relative change of the Rabi frequency across the BEC still remains below $10^{-3}$ and can, therefore, be neglected in our considerations, too.

\section{Conclusion and outlook}\label{sec:conclusion}
In this paper we have introduced an rf outcoupling method that allows us to generate a coherent and isotropic shell of atoms slowly expanding in microgravity, the so-called space atom laser. What makes our method unique is the fact that the isotropic distribution in position and momentum emerges naturally during the outcoupling process and is essentially independent of the geometry of the magnetic trap. Hence, even if the initial BEC was trapped in an elongated anisotropic trap, the outcoupled atoms form a spherically symmetric shell. 
This isotropy of the space atom laser shell is an interesting property in its own right since it would be the first time that such a large, highly symmetric matter wave could be realized experimentally.

Our model sequence is based on realistic values and shows that an experimental realization of the space atom laser is within reach of upcoming microgravity experiments such as NASA's CAL aboard the ISS. We have demonstrated that it is in principle possible to establish a spatially well-localized as well as state-selective outcoupling that leads to a large outcoupling efficiency, small expansion rates and a high isotropy of the output state. With our model sequence we have achieved average 3D-expansion velocities of around $400\,\mu\mathrm{m}/\mathrm{s}$ with a width of the velocity distribution of roughly $100\,\mu\mathrm{m}/\mathrm{s}$. Moreover, with dedicated sequences even lower velocities can be reached if at the same time one accepts either a lower number of particles transferred to the output state or a reduced isotropy. 

By scanning the Rabi frequency we have identified different regimes of operation for the space atom laser including an oscillatory behavior for large values of $\ORabi$, and a sweet spot for operation at intermediate values. We have also studied in detail the influence of a varying number of particles and fluctuations of the magnetic field on the outcoupling results, which is important for future experimental realizations of the space atom laser. 

Furthermore, we have found that the higher-order contributions to the Zeeman effect generate a small anti-trapping potential for the $F=1, \mf=0$ hyperfine sublevel which leads to an anisotropic broadening of the momentum distribution of the output state. In this context we have also discussed possible strategies to suppress this effect by changing the experimental setup. The possibility of using microwave fields to create an effective two-level system seems particularly promising and will be the subject of future studies.

With the space atom laser we open up new possibilities in matter-wave optics by enabling the generation of freely expanding shell-shaped BECs, which can be seen as a first step towards the custom-built design of matter-wave packets in microgravity. 
In fact, the expansion velocity of the outcoupled atoms is a function of the resonance position of the rf field with respect to the trapped BEC and the shape of the momentum distribution of the output state crucially depends on the value of the Rabi frequency.
This tuneability allows in principle to control the size and thickness of the shell taking into account the experimental needs. 
We expect this approach to be further developed in the future and to enable the generation of even more exotic wave functions for applications in matter-wave optics. 

Another way of realizing matter waves with shell geometries in microgravity which is related to our method is to load a BEC into a bubble-shaped trap generated by rf-dressed potentials \cite{Zobay2004,Perrin2017,Colombe2004,Lesanovsky2007}. 
In this case a strong rf coupling is required, in contrast to the weak coupling employed for the space atom laser. While our work focuses on generating freely expanding matter-wave shells with almost perfectly isotropic density distributions, rf-dressing allows to study trapped BECs with a shell geometry. 
An alternative to rf-dressed potentials are dynamically ``painted'' optical potentials \cite{Henderson2009}, which enable the generation of a wide range of different trapping geometries.
A number of interesting phenomena can be investigated in these hollow BECs, such as the creation of vortices on thin surfaces and the realization of a tuneable transition between 3D and 2D topologies \cite{Padavic2017}. 
In fact, these bubble-trapped BECs would be the 3D version of ring-trapped BECs \cite{Heathcote2008,Sherlock2011,Navez2016,Bell2016} and can only be realized in microgravity \cite{Lundblad2016} due to the sagging effect present in an earth-bound laboratory. 
When released from the bubble-shaped trap, however, the BEC does not keep its shell-like form, instead it partly collapses and expands, which results in a much broader density distribution compared with the space atom laser. 

In the same way that the generation of toroidal BECs has offered the possibility of studying many interesting physical phenomena connected with their non-trivial topology \cite{Wright2013,Ryu2013,Eckel2014}, producing BECs with a spherical geometry will provide an opportunity for investigating a wealth of new phenomena that have remained inaccessible to ground-based experiments. Furthermore, in the space atom laser these features are combined with the appealing coherence properties of atom lasers and should enable the conception of radically new realizations of matter-wave interferometry, for example schemes involving superpositions of concentrical shells expanding with different radial velocities which are eventually recombined.

\section*{Acknowledgments}
We thank D.~Aveline, N.~Bigelow, E.~Elliott, A.~Friedrich, N.~Gaaloul, E.~Giese, W.~Herr, W.~Ketterle, M.~Lee, H.~Losert, N.~Lundblad, W.~D.~Phillips, R.~Thompson, J.~Williams, N.~Yu and W.~Zeller for fruitful discussions and helpful suggestions throughout the various stages of this work.
This project was generously supported by the German Space Agency (DLR) with funds provided by the Federal Ministry for Economic Affairs and Energy (BMWi) under Grant No. 50WM1556 (QUANTUS IV).
The research of the IQ$^\mathrm{ST}$ is financially supported by the Ministry of Science, Research and Arts Baden-W\"{u}rttemberg.
W.~P.~S. is most grateful to Texas A\&M University for a Faculty Fellowship at the Hagler Institute for Advanced Study at the Texas A\&M University as well as to the Texas A\&M AgriLife Research for its support.

\appendix
\section{The Zeeman effect}\label{app:sec:general:zeeman:effect}
In this appendix we discuss the influence of the higher-order contributions of the Zeeman effect on the trapping potentials generated by spatially varying magnetic fields. We first recall the Breit-Rabi formula \cite{BergmannSchaefer1997, BreitRabi1931} and then calculate the energy difference between two hyperfine transitions as well as derive a formula for the anti-trapping potential experienced by the $\mf=0$ hyperfine state.

\subsection{The Breit-Rabi formula}\label{app:sec:Breit_Rabi}
The Breit-Rabi formula \cite{BergmannSchaefer1997, BreitRabi1931} describes the regime of intermediate magnetic field strength between the \emph{linear Zeeman effect} and the \emph{Paschen-Back effect}. According to the Breit-Rabi formula the external potential for an atom in the ground-state manifold of the D~transition with total electron angular momentum $J = 1/2$ and total nuclear angular momentum $I$ subject to a magnetic field $\vec{B}(\x,t)$ is given by
\begin{eqnarray}\label{eq:app:Breit-Rabi_full}
 V_{|J=1/2,m_J,I,m_I>}(\x,t) =& - \frac{\Delta E_{\mathrm{hfs}}}{2(2I+1)} - m\, g_I\, \mu_\mathrm{B}\, \big|\vec{B}(\x,t)\big| \nonumber \\
 & \pm \frac{\Delta E_{\mathrm{hfs}}}{2} \sqrt{1 + \frac{4 m}{2I+1} b(\x,t) + b^2(\x,t)}
\end{eqnarray}
with
\begin{equation}\label{eq:app:def:b}
 b(\x,t) = \frac{(g_J - g_I)\, \mu_\mathrm{B}\, \big|\vec{B}(\x,t)\big|}{\Delta E_{\mathrm{hfs}}} \,.
\end{equation}
Here, $\Delta E_{\mathrm{hfs}} = A_{\mathrm{hfs}}(I + 1/2)$ denotes the hyperfine splitting energy, $A_{\mathrm{hfs}}$ is the zero field hyperfine constant of the atom, $g_I$ and $g_J$ are the nuclear $g$-factor and the fine structure $g$-factor, respectively, and $\mu_\mathrm{B}$ is the Bohr magneton. In addition, $m = m_I \pm m_J$, where the plus-minus sign is chosen to coincide with the one in Eq. \eqref{eq:app:Breit-Rabi_full}.

Since in this article we focus on the $F=1$ ground-state hyperfine manifold of $^{87}$Rb, which features the quantum numbers $J = 1/2$ and $I = 3/2$, we can simplify Eq. \eqref{eq:app:Breit-Rabi_full} for this special case by taking the negative sign in front of the last term of Eq. \eqref{eq:app:Breit-Rabi_full} and by identifying $m = \mf$. As a result, we obtain the external potential 
\begin{equation}\label{eq:app:Breit-Rabi_Rb_only}
\fl V_{\mathrm{ext},\mf}(\x,t) = - \frac{A_\mathrm{hfs}}{4} - \mf\, g_I\, \mu_\mathrm{B}\, \big|\vec{B}(\x,t)\big| - A_\mathrm{hfs} \sqrt{1 + \mf\,b(\x,t) + b^2(\x,t) }
\end{equation}
for the hyperfine sublevels $\mf = -1$, $0$ and $+1$.

With the help of Eq. \eqref{eq:app:Breit-Rabi_Rb_only} we can evaluate the difference between the two transition frequencies \eqref{eq:reso:frequency:-1:0} and \eqref{eq:reso:frequency:0:+1}, which yields
\begin{equation}\label{eq:app:difference_frequencies}
 \omega_{-1,0} - \omega_{0,+1} = \frac{A_\mathrm{hfs}}{\hbar} \left[2\sqrt{1 + b^2} - \sqrt{1 - b + b^2} - \sqrt{1 + b + b^2} \right] > 0 \, ,
\end{equation}
where the inequality can be easily verified by comparing the square of the first term with the square of the sum of the second and third term within the square brackets in Eq. \eqref{eq:app:difference_frequencies}. Hence, we have proven that there is an asymmetry, reflected by Eq. \eqref{eq:transition:frequencies:not:equal}, between $\omega_{-1,0}$ and $\omega_{0,+1}$.

\subsection{Anti-trapping potential for the $\mf = 0$ hyperfine sublevel}\label{app:sec:mf0_pot}
According to Eq. \eqref{eq:app:Breit-Rabi_Rb_only} the external potential for the $\mf = 0$ hyperfine state does depend on the magnetic field $\vec{B}(\x,t)$ through the quantity $b(\x,t)$, Eq. \eqref{eq:app:def:b}, and can thus vary in space. For small magnetic fields, that is $b^2(\x,t) \ll 1$, one can expand the square root in Eq. \eqref{eq:app:Breit-Rabi_Rb_only} around $1$, which yields the potential
\begin{equation}\label{eq:app:mf0_pot_step_1}
 V_{\mathrm{ext},0}(\x,t) = - \frac{5\, A_\mathrm{hfs}}{4} - \frac{A_\mathrm{hfs} \, b^2(\x,t)}{2} + O(b^4(\x,t)) \;.
\end{equation}
By applying Eqs. \eqref{eq:app:def:b} and \eqref{eq:magnetic:field:split} one can cast Eq. \eqref{eq:app:mf0_pot_step_1} into the following form:
\begin{eqnarray}\label{eq:app:mf0_pot_step_2}
 \fl V_{\mathrm{ext},0}(\x,t) = & - \frac{5\, A_\mathrm{hfs}}{4} - \frac{(g_J - g_I)^2 \,\mu_\mathrm{B}^2}{8\,A_\mathrm{hfs}} \left[B_\mathrm{bot}^2 + 2 \,B_\mathrm{bot}\,B_\mathrm{trap}(\x,t) + B_\mathrm{trap}^2(\x,t) \right] \\
 & + O\left(|\vec{B}(\x,t)|^4\right) \;.
\end{eqnarray}
Hence, the spatial-dependent potential for the $\mf=0$ state is given up to first order in $B_\mathrm{trap}(\x,t)$ by 
\begin{equation}\label{eq:app:mf0_pot_step_3}
 V_{\mathrm{trap},0}(\x,t) \approx - \frac{(g_J - g_I)^2 \,\mu_\mathrm{B}^2}{4\,A_\mathrm{hfs}} \,B_\mathrm{bot}\,B_\mathrm{trap}(\x,t) \;.
\end{equation}
Since the right-hand side of Eq. \eqref{eq:app:mf0_pot_step_3} is negative, atoms in the $F=1,\mf=0$ hyperfine state of $^{87}$Rb are subject to a small anti-trapping potential.
In order to compare the potentials for the $\mf=0$ and the $\mf=-1$ hyperfine state one can analogously derive an relation for $\mf=-1$, which basically gives rise to Eq. \eqref{eq:lin:Zeeman:effect} for the linear Zeeman effect:
\begin{equation}\label{eq:app:linear:zeeman:effect}
 V_{\mathrm{trap},-1}(\x,t) \approx |g_F| \,\mu_\mathrm{B}\,B_\mathrm{trap}(\x,t) \;.
\end{equation}
Finally, combining Eqs. \eqref{eq:app:mf0_pot_step_3} and \eqref{eq:app:linear:zeeman:effect} yields the proportionality
\begin{equation}
 V_{\mathrm{trap},0}(\x,t) \approx - \frac{(g_J - g_I)^2 \,\mu_\mathrm{B}}{4\,|g_F|\,A_\mathrm{hfs}}\,B_\mathrm{bot}\,V_{\mathrm{trap},-1}(\x,t) 
\end{equation}
between the two potentials.

\section{Requirements for a spatially-dependent and state-selective outcoupling}\label{app:sec:consitions:sharp:and:state}
In this appendix we present two conditions that need to be satisfied in order to accurately outcouple only the atoms at the very edge of the BEC from the trapped state to the untrapped state. For this purpose we (i) discuss how the transition resonance can be made sharp and spatially well localized, and (ii) address the challenge of selectively driving a single transition within a multi-level system.

\subsection{Sharp resonance condition}\label{app:sharp:resonance}
Since the transition frequencies \eqref{eq:reso:frequency:-1:0} and \eqref{eq:reso:frequency:0:+1} are spatially dependent the resonance conditions \eqref{eq:reso:condition:-1:0} and \eqref{eq:reso:condition:0:+1} are in general only fulfilled on a two-dimensional surface in three-dimensional space. However, in analogy to Rabi oscillations the transition between two states can be driven even off resonance depending on the detuning and the coupling strength. 
Based on the formalism for Rabi oscillations we derive a condition which, if fulfilled, ensures that the resonance for the transition from one hyperfine sublevel to the next one is sharp enough so that only a fraction of the BEC is outcoupled at a given instant. We want to stress that this procedure oversimplifies the dynamics of our system and therefore yields only a rough estimate for the parameter range in which a sharp resonance can be expected. Nevertheless, our condition is still a valuable starting point for exploring the dynamics of the space atom laser. 

When we consider the transition from the trapped state to the untrapped one the amplitude of the corresponding Rabi oscillations is given by
\begin{equation}\label{eq:app:transition_amplitude_def}
 \sigma_{0,-1}(\x,t) = \frac{\ORabi^2}{\ORabi^2 + [\omega_{-1,0}(\x,t) - \wrf(t)]^2} \,.
\end{equation}
For a sharp resonance we require that $\sigma_{0,-1} \ll 1$ at the center of the trap when the resonance condition is fulfilled at the very edge of the condensate. In this way, the transition from one state to the other only takes place at the edge of the BEC while the central region is not outcoupled at all. In this case the detuning $\omega_{-1,0}(\x,t) - \wrf(t) \approx - \mu_{-1}/\hbar$ as illustrated in Fig.~\ref{fig:energy_levels}. Consequently, we obtain the inequality
\begin{equation}\label{eq:app:condition:sharp:resonance:step2}
 \frac{1}{1 + \mu_{-1}^2/(\hbar\ORabi)^2} \ll 1 \;,
\end{equation}
which clearly is fulfilled if the condition
\begin{equation}\label{eq:app:condition:sharp:resonance:final}
 \mu_{-1}^2 \gg (\hbar\ORabi)^2 
\end{equation}
holds true. Hence, for a given chemical potential $\mu_{-1}$, which characterizes the initial state of the BEC, Eq. \eqref{eq:app:condition:sharp:resonance:final} sets an upper bound on the Rabi frequency $\ORabi$ for spatially-selective outcoupling. 

\subsection{State-selective outcoupling}\label{app:state:selective}
In addition to having a sharp resonance guaranteed by condition~\eqref{eq:app:condition:sharp:resonance:final}, we want to make sure that the atoms are only coupled from the trapped state to the untrapped state and that the coupling to the anti-trapped state is suppressed. This can be achieved by raising the lower bound in Eq. \eqref{eq:app:difference_frequencies} and demanding that
\begin{equation}\label{eq:app:state_selective_step1}
 \omega_{-1,0}(\x,t) - \omega_{0,+1}(\x,t) \gg \frac{\mu_{-1}}{\hbar} \,.
\end{equation}
Again the chemical potential $\mu_{-1}$ is the appropriate bound for this inequality because the combination of Eq. \eqref{eq:app:condition:sharp:resonance:final} and Eq. \eqref{eq:app:state_selective_step1} ensures that the detuning for the transition from the untrapped to the anti-trapped state is large enough to suppress this transition completely. With the help of the first equality sign in Eq. \eqref{eq:app:difference_frequencies} we can rewrite Eq. \eqref{eq:app:state_selective_step1} as
\begin{equation}\label{eq:app:state_selective_step2}
\fl 2\sqrt{1 + b^2(\x,t)} - \sqrt{1 - b(\x,t) + b^2(\x,t)} - \sqrt{1 + b(\x,t) + b^2(\x,t)} \gg \frac{\mu_{-1}}{A_\mathrm{hfs}} \,.
\end{equation}

Since for typical magnetic traps the magnetic field strengths used are quite small we have $b(\x,t) \ll 1$ in the region where the trapped atoms are actually located. Hence, by expanding the square roots in Eq. \eqref{eq:app:state_selective_step2} around one up to second order in $b$, we obtain
\begin{equation}\label{eq:app:state_selective_step3}
 \frac{1}{4}b^2 + O(b^3) \gg \frac{\mu_{-1}}{A_\mathrm{hfs}} \,.
\end{equation}
By taking into account the definition of $b(\x,t)$ in Eq. \eqref{eq:app:def:b} and the fact that the magnitude of the magnetic field $\vec{B}(\x,t)$ takes it minimum value at the center of the trap, we finally get the condition
\begin{equation}\label{eq:app:state_selective_final}
 B_\mathrm{bot}^2 \gg \frac{16 \,\mu_{-1} \,A_\mathrm{hfs}}{(g_J - g_I)^2 \,\mu_\mathrm{B}^2} 
\end{equation}
for the magnetic field strength $B_\mathrm{bot}$ at the trap bottom which ensures a state-selective outcoupling process.

\section*{References}
\bibliographystyle{nature}
\bibliography{bibfile_space_atom_laser}

\begin{thebibliography}{10}

\bibitem{Anderson1995}
Anderson, M.~H., Ensher, J.~R., Matthews, M.~R., Wieman, C.~E., and Cornell,
  E.~A.
\newblock {\em Science}{ \bf 269}, 198--201 (1995).

\bibitem{Davis1995}
Davis, K.~B., Mewes, M.~O., Andrews, M.~R., van Druten, N.~J., Durfee, D.~S.,
  Kurn, D.~M., and Ketterle, W.
\newblock {\em Phys. Rev. Lett.}{ \bf 75}, 3969--3973 (1995).

\bibitem{Mewes1997}
Mewes, M.-O., Andrews, M.~R., Kurn, D.~M., Durfee, D.~S., Townsend, C.~G., and
  Ketterle, W.
\newblock {\em Phys. Rev. Lett.}{ \bf 78}, 582--585 (1997).

\bibitem{Anderson1998}
Anderson, B.~P. and Kasevich, M.~A.
\newblock {\em Science}{ \bf 282}, 1686--1689 (1998).

\bibitem{Hagley1999}
Hagley, E.~W., Deng, L., Kozuma, M., Wen, J., Helmerson, K., Rolston, S.~L.,
  and Phillips, W.~D.
\newblock {\em Science}{ \bf 283}, 1706--1709 (1999).

\bibitem{Bloch1999}
Bloch, I., H\"ansch, T.~W., and Esslinger, T.
\newblock {\em Phys. Rev. Lett.}{ \bf 82}, 3008--3011 (1999).

\bibitem{Ballagh1997}
Ballagh, R.~J., Burnett, K., and Scott, T.~F.
\newblock {\em Phys. Rev. Lett.}{ \bf 78}, 1607--1611 (1997).

\bibitem{Naraschewski1997}
Naraschewski, M., Schenzle, A., and Wallis, H.
\newblock {\em Phys. Rev. A}{ \bf 56}, 603--606 (1997).

\bibitem{Steck1997}
Steck, H., Naraschewski, M., and Wallis, H.
\newblock {\em Phys. Rev. Lett.}{ \bf 80}, 1--5 (1998).

\bibitem{Kneer1998}
Kneer, B., Wong, T., Vogel, K., Schleich, W.~P., and Walls, D.~F.
\newblock {\em Phys. Rev. A}{ \bf 58}, 4841--4853 (1998).

\bibitem{Kramer2002}
Kramer, T., Bracher, C., and Kleber, M.
\newblock {\em Journal of Physics A: Mathematical and General}{ \bf 35}(40),
  8361 (2002).

\bibitem{Bracher2003}
Bracher, C., Kramer, T., and Kleber, M.
\newblock {\em Phys. Rev. A}{ \bf 67}, 043601 (2003).

\bibitem{Kramer2006}
Kramer, T. and Rodr\'{\i}guez, M.
\newblock {\em Phys. Rev. A}{ \bf 74}, 013611 (2006).

\bibitem{Johnsson2007}
Johnsson, M., Haine, S., Hope, J., Robins, N., Figl, C., Jeppesen, M., Dugu\'e,
  J., and Close, J.
\newblock {\em Phys. Rev. A}{ \bf 75}, 043618 (2007).

\bibitem{Robins2013}
Robins, N., Altin, P., Debs, J., and Close, J.
\newblock {\em Physics Reports}{ \bf 529}(3), 265 -- 296 (2013).

\bibitem{Bloch2000}
Bloch, I., H\"ansch, T.~W., and Esslinger, T.
\newblock {\em Nature}{ \bf 403}, 166--170 (2000).

\bibitem{Koehl2001}
K\"ohl, M., H\"ansch, T.~W., and Esslinger, T.
\newblock {\em Phys. Rev. Lett.}{ \bf 87}, 160404 (2001).

\bibitem{Koehl2005}
K\"ohl, M., Busch, T., M\o{}lmer, K., H\"ansch, T.~W., and Esslinger, T.
\newblock {\em Phys. Rev. A}{ \bf 72}, 063618 (2005).

\bibitem{Riou2006}
Riou, J.-F., Guerin, W., Le~Coq, Y., Fauquembergue, M., Josse, V., Bouyer, P.,
  and Aspect, A.
\newblock {\em Phys. Rev. Lett.}{ \bf 96}, 070404 (2006).

\bibitem{Fabre2011}
Fabre, C.~M., Cheiney, P., Gattobigio, G.~L., Vermersch, F., Faure, S.,
  Mathevet, R., Lahaye, T., and Gu\'ery-Odelin, D.
\newblock {\em Phys. Rev. Lett.}{ \bf 107}, 230401 (2011).

\bibitem{Gattobigio2012}
Gattobigio, G.~L., Couvert, A., Reinaudi, G., Georgeot, B., and Gu\'ery-Odelin,
  D.
\newblock {\em Phys. Rev. Lett.}{ \bf 109}, 030403 (2012).

\bibitem{Bolpasi2014}
Bolpasi, V., Efremidis, N.~K., Morrissey, M.~J., Condylis, P.~C., Sahagun, D.,
  Baker, M., and von Klitzing, W.
\newblock {\em New Journal of Physics}{ \bf 16}(3), 033036 (2014).

\bibitem{KleineBuening2010}
Kleine~B\"uning, G., Will, J., Ertmer, W., Klempt, C., and Arlt, J.
\newblock {\em Applied Physics B}{ \bf 100}(1), 117--123 (2010).

\bibitem{Guerin2006}
Guerin, W., Riou, J.-F., Gaebler, J.~P., Josse, V., Bouyer, P., and Aspect, A.
\newblock {\em Phys. Rev. Lett.}{ \bf 97}, 200402 (2006).

\bibitem{Couvert2008}
Couvert, A., Jeppesen, M., Kawalec, T., Reinaudi, G., Mathevet, R., and
  Gu\'ery-Odelin, D.
\newblock {\em EPL (Europhysics Letters)}{ \bf 83}(5), 50001 (2008).

\bibitem{Bernard2011}
Bernard, A., Guerin, W., Billy, J., Jendrzejewski, F., Cheinet, P., Aspect, A.,
  Josse, V., and Bouyer, P.
\newblock {\em New Journal of Physics}{ \bf 13}(6), 065015 (2011).

\bibitem{VanZoest2010}
van Zoest, T., Gaaloul, N., Singh, Y., Ahlers, H., Herr, W., Seidel, S.~T.,
  Ertmer, W., Rasel, E., Eckart, M., Kajari, E., Arnold, S., Nandi, G.,
  Schleich, W.~P., Walser, R., Vogel, A., Sengstock, K., Bongs, K.,
  Lewoczko-Adamczyk, W., Schiemangk, M., Schuldt, T., Peters, A., K{\"o}nemann,
  T., M{\"u}ntinga, H., L{\"a}mmerzahl, C., Dittus, H., Steinmetz, T.,
  H{\"a}nsch, T.~W., and Reichel, J.
\newblock {\em Science}{ \bf 328}, 1540--1543 (2010).

\bibitem{Muentinga2013}
M\"untinga, H., Ahlers, H., Krutzik, M., Wenzlawski, A., Arnold, S., Becker,
  D., Bongs, K., Dittus, H., Duncker, H., Gaaloul, N., Gherasim, C., Giese, E.,
  Grzeschik, C., H\"ansch, T.~W., Hellmig, O., Herr, W., Herrmann, S., Kajari,
  E., Kleinert, S., L\"ammerzahl, C., Lewoczko-Adamczyk, W., Malcolm, J.,
  Meyer, N., Nolte, R., Peters, A., Popp, M., Reichel, J., Roura, A., Rudolph,
  J., Schiemangk, M., Schneider, M., Seidel, S.~T., Sengstock, K., Tamma, V.,
  Valenzuela, T., Vogel, A., Walser, R., Wendrich, T., Windpassinger, P.,
  Zeller, W., van Zoest, T., Ertmer, W., Schleich, W.~P., and Rasel, E.~M.
\newblock {\em Phys. Rev. Lett.}{ \bf 110}, 093602 (2013).

\bibitem{Becker2018}
Becker, D., Lachmann, M.~D., Seidel, S.~T., Ahlers, H., Dinkelaker, A.~N.,
  Grosse, J., Hellmig, O., M\"untinga, H., Schkolnik, V., Wendrich, T.,
  Wenzlawski, A., Weps, B., Corgier, R., Franz, T., Gaaloul, N., Herr, W.,
  L\"udtke, D., Popp, M., Amri, S., Duncker, H., Erbe, M., Kohfeldt, A.,
  Kubelka-Lange, A., Braxmaier, C., Charron, E., Ertmer, W., Krutzik, M.,
  L\"ammerzahl, C., Peters, A., Schleich, W.~P., Sengstock, K., Walser, R.,
  Wicht, A., Windpassinger, P., and Rasel, E.~M.
\newblock {\em Nature}{ \bf 562}, 391--395 (2018).

\bibitem{Martin1999}
Martin, J.~L., McKenzie, C.~R., Thomas, N.~R., Sharpe, J.~C., Warrington,
  D.~M., Manson, P.~J., Sandle, W.~J., and Wilson, A.~C.
\newblock {\em Journal of Physics B: Atomic, Molecular and Optical Physics}{
  \bf 32}(12), 3065 (1999).

\bibitem{LeCoq2001}
Le~Coq, Y., Thywissen, J.~H., Rangwala, S.~A., Gerbier, F., Richard, S.,
  Delannoy, G., Bouyer, P., and Aspect, A.
\newblock {\em Phys. Rev. Lett.}{ \bf 87}, 170403 (2001).

\bibitem{Oettl2005}
\"Ottl, A., Ritter, S., K\"ohl, M., and Esslinger, T.
\newblock {\em Phys. Rev. Lett.}{ \bf 95}, 090404 (2005).

\bibitem{Oettl2006}
\"Ottl, A., Ritter, S., K\"ohl, M., and Esslinger, T.
\newblock {\em Review of Scientific Instruments}{ \bf 77}(6), 063118 (2006).

\bibitem{Robins2006}
Robins, N.~P., Figl, C., Haine, S.~A., Morrison, A.~K., Jeppesen, M., Hope,
  J.~J., and Close, J.~D.
\newblock {\em Phys. Rev. Lett.}{ \bf 96}, 140403 (2006).

\bibitem{Debs2009}
Debs, J.~E., D\"{o}ring, D., Robins, N.~P., Figl, C., Altin, P.~A., and Close,
  J.~D.
\newblock {\em Opt. Express}{ \bf 17}(4), 2319--2325 (2009).

\bibitem{Billy2007}
{Billy, J.}, {Josse, V.}, {Zuo, Z.}, {Guerin, W.}, {Aspect, A.}, and {Bouyer,
  P.}
\newblock {\em Ann. Phys. Fr.}{ \bf 32}(2-3), 17--24 (2007).

\bibitem{Elliott2018}
Elliott, E.~R., Krutzik, M.~C., Williams, J.~R., Thompson, R.~J., and Aveline,
  D.~C.
\newblock {\em npj Microgravity}{ \bf 4}, 16 (2018).

\bibitem{Dalfovo1999}
Dalfovo, F., Giorgini, S., Pitaevskii, L.~P., and Stringari, S.
\newblock {\em Rev. Mod. Phys.}{ \bf 71}, 463--512 (1999).

\bibitem{StamperKurn2013}
Stamper-Kurn, D.~M. and Ueda, M.
\newblock {\em Rev. Mod. Phys.}{ \bf 85}, 1191--1244 (2013).

\bibitem{BergmannSchaefer1997}
Bergmann, L. and Schaefer, C.
\newblock {\em Constituents of Matter - Atoms, Molecules, Nuclei and
  Particles}.
\newblock Walter de Gruyter, Berlin, New York,  (1997).

\bibitem{BreitRabi1931}
Breit, G. and Rabi, I.~I.
\newblock {\em Phys. Rev.}{ \bf 38}, 2082--2083 (1931).

\bibitem{Schleich2001}
Schleich, W.~P.
\newblock {\em Quantum Optics in Phase Space}.
\newblock Wiley-VCH, Berlin,  (2001).

\bibitem{Nandi2007}
Nandi, G., Walser, R., Kajari, E., and Schleich, W.~P.
\newblock {\em Phys. Rev. A}{ \bf 76}, 063617 (2007).

\bibitem{Meister2017}
Meister, M., Arnold, S., Moll, D., Eckart, M., Kajari, E., Efremov, M.~A.,
  Walser, R., and Schleich, W.~P.
\newblock In {\em Advances In Atomic, Molecular, and Optical Physics,
  }Arimondo, E., Lin, C.~C., and Yelin, S.~F., editors, volume~66,  375 -- 438.
  Academic Press (2017).

\bibitem{CAL_Homepage}
More information about {CAL} can be found on the mission homepage:
  https://coldatomlab.jpl.nasa.gov.

\bibitem{Auer2001}
Auer, J., Krotscheck, E., and Chin, S.~A.
\newblock {\em The Journal of Chemical Physics}{ \bf 115}(15), 6841--6846
  (2001).

\bibitem{Press2007}
Press, W.~H., Teukolsky, S.~A., Vetterling, W.~T., and Flannery, B.~P.
\newblock {\em Numerical Recipes - The Art of Scientific Computing}.
\newblock Cambridge University Press, Cambridge, third edition,  (2007).

\bibitem{Neuhauser1989}
Neuhauser, D. and Baer, M.
\newblock {\em The Journal of Chemical Physics}{ \bf 90}(8), 4351--4355 (1989).

\bibitem{McDonald2013}
McDonald, G.~D., Kuhn, C. C.~N., Bennetts, S., Debs, J.~E., Hardman, K.~S.,
  Johnsson, M., Close, J.~D., and Robins, N.~P.
\newblock {\em Phys. Rev. A}{ \bf 88}, 053620 (2013).

\bibitem{Kovachy2015}
Kovachy, T., Hogan, J.~M., Sugarbaker, A., Dickerson, S.~M., Donnelly, C.~A.,
  Overstreet, C., and Kasevich, M.~A.
\newblock {\em Phys. Rev. Lett.}{ \bf 114}, 143004 (2015).

\bibitem{Zobay2004}
Zobay, O. and Garraway, B.~M.
\newblock {\em Phys. Rev. A}{ \bf 69}, 023605 (2004).

\bibitem{Perrin2017}
Perrin, H. and Garraway, B.~M.
\newblock In {\em Advances In Atomic, Molecular, and Optical Physics,
  }E.~Arimondo, C. C.~L. and Yelin, S.~F., editors, volume~66,  181 -- 262.
  Academic Press (2017).

\bibitem{Colombe2004}
Colombe, Y., Knyazchyan, E., Morizot, O., Mercier, B., Lorent, V., and Perrin,
  H.
\newblock {\em EPL (Europhysics Letters)}{ \bf 67}(4), 593 (2004).

\bibitem{Lesanovsky2007}
Lesanovsky, I. and von Klitzing, W.
\newblock {\em Phys. Rev. Lett.}{ \bf 99}, 083001 (2007).

\bibitem{Robins2005}
Robins, N.~P., Morrison, A.~K., Hope, J.~J., and Close, J.~D.
\newblock {\em Phys. Rev. A}{ \bf 72}, 031606 (2005).

\bibitem{Debs2010}
Debs, J.~E., D\"oring, D., Altin, P.~A., Figl, C., Dugu\'e, J., Jeppesen, M.,
  Schultz, J.~T., Robins, N.~P., and Close, J.~D.
\newblock {\em Phys. Rev. A}{ \bf 81}, 013618 (2010).

\bibitem{Henderson2009}
Henderson, K., Ryu, C., MacCormick, C., and Boshier, M.~G.
\newblock {\em New Journal of Physics}{ \bf 11}(4), 043030 (2009).

\bibitem{Padavic2017}
Padavi\'c, K., Sun, K., Lannert, C., and Vishveshwara, S.
\newblock {\em EPL (Europhysics Letters)}{ \bf 120}(2), 20004 (2017).

\bibitem{Heathcote2008}
Heathcote, W.~H., Nugent, E., Sheard, B.~T., and Foot, C.~J.
\newblock {\em New Journal of Physics}{ \bf 10}(4), 043012 (2008).

\bibitem{Sherlock2011}
Sherlock, B.~E., Gildemeister, M., Owen, E., Nugent, E., and Foot, C.~J.
\newblock {\em Phys. Rev. A}{ \bf 83}, 043408 (2011).

\bibitem{Navez2016}
Navez, P., Pandey, S., Mas, H., Poulios, K., Fernholz, T., and von Klitzing, W.
\newblock {\em New Journal of Physics}{ \bf 18}(7), 075014 (2016).

\bibitem{Bell2016}
Bell, T.~A., Glidden, J. A.~P., Humbert, L., Bromley, M. W.~J., Haine, S.~A.,
  Davis, M.~J., Neely, T.~W., Baker, M.~A., and Rubinsztein-Dunlop, H.
\newblock {\em New Journal of Physics}{ \bf 18}(3), 035003 (2016).

\bibitem{Lundblad2016}
Lundblad, N., Jarvis, T., Paseltiner, D., and Lannert, C.
\newblock In {\em APS Division of Atomic and Molecular Physics Meeting 2016},
  K1.118,  (2016).

\bibitem{Wright2013}
Wright, K.~C., Blakestad, R.~B., Lobb, C.~J., Phillips, W.~D., and Campbell,
  G.~K.
\newblock {\em Phys. Rev. Lett.}{ \bf 110}, 025302 (2013).

\bibitem{Ryu2013}
Ryu, C., Blackburn, P.~W., Blinova, A.~A., and Boshier, M.~G.
\newblock {\em Phys. Rev. Lett.}{ \bf 111}, 205301 (2013).

\bibitem{Eckel2014}
Eckel, S., Lee, J.~G., Jendrzejewski, F., Murray, N., Clark, C.~W., Lobb,
  C.~J., Phillips, W.~D., Edwards, M., and Campbell, G.~K.
\newblock {\em Nature}{ \bf 506}, 200 (2014).

\end{thebibliography}

\end{document}